\documentclass[acmsmall]{acmart}

\usepackage{float}
\usepackage{subfigure}
\usepackage[normalem]{ulem}
\usepackage{tabularray}
\AtBeginDocument{%
  \providecommand\BibTeX{{%
    \normalfont B\kern-0.5em{\scshape i\kern-0.25em b}\kern-0.8em\TeX}}}
\usepackage{tabularx}
\usepackage{bbding}
\usepackage{array} 
\usepackage{graphicx}
\usepackage{multirow}
\usepackage{placeins}
\usepackage{float} 
\usepackage{geometry}
\usepackage{subfigure}
\usepackage{amsmath}
\usepackage{makecell}
\usepackage{float}
\usepackage{color}
\usepackage{booktabs}
\usepackage{caption}
\usepackage{tabu}
\usepackage{hyperref}
\usepackage{booktabs}
\usepackage{array}
\usepackage{graphicx} 
\usepackage{longtable}
\usepackage{rotating}
\usepackage{ragged2e}
\newcolumntype{P}[1]{>{\RaggedRight\arraybackslash}p{#1}}
\usepackage[normalem]{ulem}
\usepackage{xcolor}

\newcommand{\rt}[1]{}
\usepackage[commandnameprefix=always]{changes}

\setcopyright{cc}
\setcctype{by}
\acmDOI{10.1145/3748600}
\acmYear{2025}
\acmJournal{PACMHCI}
\acmVolume{9}
\acmNumber{6}
\acmArticle{GAMES005}
\acmMonth{10}
\received{2025-02-19}
\received[accepted]{2025-07-24}

\begin{document}
\title[Breaking the News]{Breaking the News: Taking the Roles of Influencer vs. Journalist in a LLM-Based Game for Raising Misinformation Awareness}

%\maketitle

\author{Huiyun Tang}
\orcid{0009-0001-6499-4716}
\authornote{These authors contributed equally to this research.}
\affiliation{%
  \institution{University of Luxembourg}
  \city{Esch-sur-Alzette}
  \country{Luxembourg}}
\email{huiyun.tang@uni.lu}

\author{Songqi Sun}
\orcid{0009-0001-5320-5515}
\authornotemark[1]
\affiliation{%
  \institution{University College London}
\city{London}
\country{United Kingdom}}
\email{songqi.sun.22@ucl.ac.uk}

\author{Kexin Nie}
\orcid{0009-0002-9190-092X}
\affiliation{%
  \institution{The University of Sydney}
\city{Sydney}
\country{Australia}}
\email{knie0519@uni.sydney.edu.au}

\author{Ang Li}
\orcid{0009-0006-4228-1092}
\affiliation{%
  \institution{Uppsala University}
\city{Uppsala}
\country{Sweden}}
\email{ang.li.4299@student.uu.se}

\author{Anastasia Sergeeva}
\orcid{0000-0003-3701-3123}
\affiliation{%
  \institution{University of Luxembourg}
  \city{Esch-sur-Alzette}
  \country{Luxembourg}}
\email{anastasia.sergeeva@uni.lu}

%do not change.
\author{Ray LC}
\authornote{Correspondences should be addressed to LC@raylc.org.}
\orcid{0000-0001-7310-8790}
\affiliation{
\institution{City University of Hong Kong}
\city{Hong Kong}
\country{China}}
\email{LC@raylc.org}

\renewcommand{\shortauthors}{Huiyun Tang, Songqi Sun, Kexin Nie, Ang Li, Anastasia Sergeeva, Ray LC}

\begin{teaserfigure}
    \centering
    \includegraphics[width=0.9\linewidth]{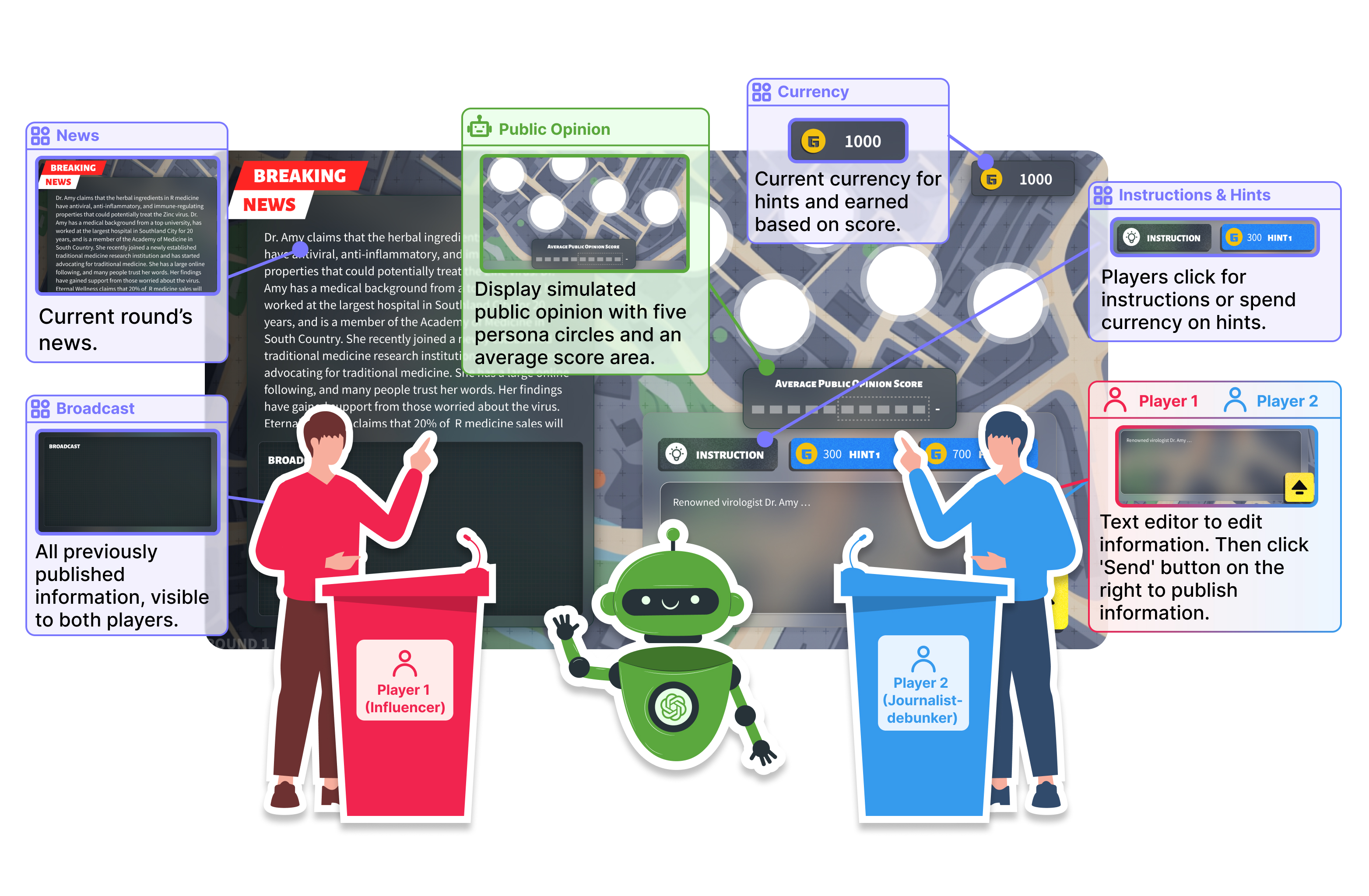}
    \caption{\textit{Breaking the News} is an online player-versus-player (PvP) game where players generate or debunk misinformation to win the trust of public opinion, represented by five LLM-driven personas.}
    \label{fig:enter-label}
\end{teaserfigure}

\begin{abstract}
%(150 words version)

% Clean version
\rt{{Informing the public about online misinformation is challenging due to unclear sources and a lack of transparency in its spread and debunking.}}

%edited by RLC 2025-07-23
Effectively mitigating online misinformation requires understanding of their mechanisms and learning of practical skills for identification and counteraction. Serious games may serve as tools for combating misinformation, teaching players to recognize common misinformation tactics, and improving their skills of discernment. However, current interventions are designed as single-player, choice-based games, which present players with limited predefined choices. Such restrictions reduce replayability and may lead to an overly simplistic understanding of misinformation and how to debunk them. This study seeks to empower people to understand opinion-influencing and misinformation-debunking processes. We created a Player vs. Player (PvP) game in which participants attempt to generate or debunk misinformation to convince the public opinion represented by LLM. Using a within-subjects mixed-methods study design (N=47), we found that this game significantly raised participants' media literacy and improved their ability to identify misinformation. Qualitative analyses revealed how participants' use of debunking and content creation strategies deepened their understanding of misinformation. This work shows the potential for illuminating  contrasting viewpoints of social issues by LLM-based mechanics in PvP games.

\end{abstract}

%%
%% The code below is copied from, generated by the tool at http://dl.acm.org/ccs.cfm.

%%
%% Keywords.

\maketitle

\section{Introduction}\label{sec:Introduction}
%editing for conciseness and coherence by RLC 2025-07-24.

%motivation
The prevalence of misinformation on social media is a growing global concern. Misinformation threatens the maintenance of trust in social agendas like vaccine and health policies \cite{do2022infodemics}, incites violence and harassment \cite{cdtFromFellows}, undermines democratic processes \cite{bovet2019influence}, and harms individual and societal well-being \cite{verma2022examining}. For example, people were even persuaded to take ineffective treatments like alcohol-based cleaning products and anti-parasitic drugs for Covid-19. Countermeasures against misinformation spread consist of two forms: preemptive intervention (``prebunking'') and reactive intervention (``debunking'')\cite{ecker2022psychological}. The latter involves correcting misinformation after it has been encountered. and using fact-checking to dispute factual inaccuracies. However, the lasting effects of misinformation make it challenging to mitigate its influence once people have been exposed \cite{roets2017fake}, and fact-checking efforts are limited in scale and reach \cite{roozenbeek2020prebunking}. Prebunking, on the other hand, works to build attitudinal inoculation, enabling people to identify and resist manipulative messaging. This approach equips individuals to manage misinformation they encounter in the real world \cite{prebin2024} based on educational measures, including games \cite{cook2023cranky,fu_cracking_2025}.

Game-based prebunking interventions engage users in simulated misinformation scenarios, allowing them to actively practice identifying and countering deceptive content\cite{kiili2024tackling}. On one hand, in games like ``Bad News,''\cite{roozenbeek2019fake} ``Harmony Square,''\cite{roozenbeek2020breaking} ``Go Viral!''\cite{camCambridgeGame} and ``Trustme!''\cite{yang2021can}, players adopt the role of a misinformation producer whose task is to create and spread misinformation as efficiently as possible. In other games, like ``MAthE''\cite{katsaounidou2019mathe} and ``Escape Fake''\cite{paraschivoiu2021escape}, players acting as fact-checkers assessing the validity of information. The choice-based formats of these games can limit replayability, requiring little cognitive effort from the player, who is presented with limited numbers of pre-generated options that reduce the involvement of players. These games are also designed for single players without utilizing multiplayer mechanics that can enhance motivation through social play\cite{lepper2021intrinsic}. Instead, collaborative and competitive gameplay can enhance the effectiveness of serious game interventions\cite{cagiltay2015effect}.
%and increase player motivation \cite{cagiltay2015effect,bellotti2010designing}.

%GAMES: 2 problems:
%1. not natural, not realistic, deterministic -> LLM solution.
%2. engagement and attitude -> two-player PvP (one against another goal).

To address the challenges of limited interactions and deterministic game paths, we aim to foster open-ended exploration and engagement through Player versus Player (PvP) mechanics. This provides players the ability to not only choose from preselected options, but also to actively create content and implement their own strategies for disproving or creating misinformation as a training ground for countering real-life misinformation.

Recent work in Large language models (LLMs) provide new interaction possibilities for in-game natural dialogue \cite{zhou2024eternagram}, generating narratives\cite{park2023generative}, non-playable characters (NPCs) interactions \cite{ashby2023personalized,uludaugli2023non,gao2024humanizing}, and role-playing scenarios\cite{xu2023exploring}. 
LLMs can be prompted to impersonate specific characters to provide appropriate dialogue \cite{zhou_eternagram_2024-1,zhou_eternagram_2025,zhang_can_2025}. Integrating LLMs into creative processes in the interaction can increase involvement and critical engagement\cite{yang_ai_2022,han_when_2024,zeng_ronaldos_2025}. As the effectiveness of many inoculation interventions tends to diminish over time, developing a replayable game that reinforces players' cognitive ``resilience'' against misinformation is an essential\cite{wells2024doomscroll}. Instead of selecting from predefined inputs, applying LLM to interactions enables users to put their input into the model and receive individual feedback tailored to this input.

Inspired by previous misinformation game interventions, we developed a PvP game called \textit{Breaking the News}, where players are assigned either the role of a misinformation creator (``Influencer'' in the game) or a counteractor against misinformation (``Journalist'' in the game). The Influencer creates misinformation posts in a social media-like environment, while the Journalist seeks to counter these messages by presenting compelling arguments. LLMs are used to represent public opinion in the ``country'' where the game events take place. The goal for both players is to earn the trust of the citizens and convince them to believe the information they present. In this paper we aim to answer the following research questions:

\textbf{RQ1:}  
How may we empower users to recognize and understand the processes of misinformation generation and misinformation debunking?

\textbf{RQ2:} 
What behaviors do players exhibit in response to game mechanics and opponent tactics?

In this paper, we present the design and evaluation of the PvP game. We conducted a mixed-method study with 47 participants, using within-subjects design and pre- and post-surveys for repeated measures. Our findings suggest that through gameplay, participants improved their ability to reflect on instances of misinformation, raised their levels of media literacy, expanded their repertoire of strategies applied to countering misinformation, and improved their discernment abilities. This study provides insight into applications of LLMs in interactive PvP mechanics for media literacy. We offer practical insights for the design of serious games aimed at combating misinformation in real-world contexts.

\section{Related Work}\label{sec:Background}
%edited by RLC for conciseness, coherence 2025-07-24.

\subsection{Characteristics of Misinformation}
\label{Characteristics of Misinformation}

The term ``misinformation'' is often used to including ``fake news'', falsehoods, malicious rumours, and conspiracy theories. Some scholars distinguish between misinformation and disinformation, with the latter referring to information deliberately crafted and spread with the intent to deceive or cause harm\cite{guess2020misinformation}. Since it is often difficult to prove intent, we use ``misinformation'' as an umbrella term for diverse forms of false information\cite{southwell2019misinformation}. 

Studies have sought to identify the key characteristics that distinguish misinformation from well-sourced, authentic information. One work analyzed the writing styles of fake information versus real news. They found that texts which can be characterized as fake news typically feature longer headlines, simpler word choices, and greater use of proper nouns and verb phrases\cite{horne2017just}. Misinformation and authentic information appear to differ in how the former can be created with the intention of triggering emotions such as fear, anxiety, or sympathy\cite{choudhary2021linguistic} using opinionated wording \cite{porat2019content}.
Source credibility also differ between authentic information and misinformation. Authentic information is shared by credible sources: reputable websites, mainstream media outlets, professional news organizations, and official publications \cite{zhang2020overview}. In contrast, fake news often originates from sources designed to generate revenue. To attract clicks, these stories use unverified quotes, inflammatory narratives, and misleading images \cite{molina2021fake}.

However, not all less reliable sources are perceived as equally untrustworthy. People often gravitate toward partisan sources that align with their political ideology, leading to variations in news consumption across the political spectrum \cite{faris2017partisanship}. This selective consumption reinforces trust in these sources, even when they may be classified as biased or unreliable \cite{pennycook2019fighting}.
Compared to authentic news, misinformation tends to develop in line with a broad dynamic pattern. One study traced the life-cycle of high-profile political rumors on Twitter over a 13 month period and found that false rumors tend to reemerge, becoming more extreme over time\cite{shin2018diffusion}. Since people are more likely to trust information that they see more often, consumers are led to believe increasingly extreme misinformation that are hard to debunk\cite{yousif2019illusion}.

\subsection{Media Literacy as Protection Against Misinformation}
Media literacy is defined as the ability to ``access, analyze, and produce information for specific outcomes'' \cite{aufderheide2018media}. Citizens often need to learn skills that can protect societies against misinformation \cite{potter2010state}. One work proposed a four-component model of media literacy, including technical competency, privacy protection, social literacy, and information awareness, the latter defined as the ability to discern between truthful and false information on social media\cite{tandoc2021developing}. Chen et al. proposed dividing the skills into critical and functional domains, as well as consuming and producing content. This critical skills often need to be developed for reflecting on the content, recognizing the motives behind publications, and creating content that includes the author's perspective as forms of social influence \cite{chenwu2011unpacking,lc_designing_2021,lc_designing_2022,song_climate_2021,song_drizzle_2022}.

%There is a distinction between ``news literacy`` \cite{malik2013challenges} and ``information literacy`` \cite{jones2021does}. 
Media literacy education has recently shifted from a protectionist position to an empowering paradigm, where people were encouraged to critically engage with media and develop skills to interpret its effects \cite{hobbs2009past}. In this paradigm, interaction with misinformation can also be considered as having educational power if it teaches the person to understand its effects. This has resulted in recent game-based interventions that use inoculation theory to enact media literacy education. Inoculation theory suggests that exposure to a weaker version of misinformation can help to develop stronger protection against future exposure \cite{grace2023examining}.

\subsection{Serious Games in Media Literacy Domain}
Game-based learning and gamification, which integrate gaming elements into education, have been studied for their ability to engage learners and facilitate active, experiential learning. These approaches provide interactive feedback and contextual problem-solving, supported by theories of effective learning \cite{tobias2014game}. Serious games or applied games is defined as games primarily aim at educational value of engagement and competition. Serious games may combine educational content with playful mechanics and game narrative for conveying educational messages interactively\cite{de2017linking,lc_chikyuchi_2022,lc_machine_2020}, often using speculative elements to push players into game worlds with metaphorical practical social good applications\cite{gong_if_2025,qian_virtual_2025}.

Serious games have begun to address misinformation education by aiming to improve media literacy \cite{roozenbeek2019fake}.
These games pit players into either creators and fact-checkers. The misinformation creator's objective is to create and spread misinformation. For example, in Bad News \cite{roozenbeek2019fake}, Harmony Square \cite{roozenbeek2020breaking}, Cat Park \cite{Gusmanson.nl_2022} and ChamberBreaker \cite{jeon2021chamberbreaker}, players are tasked with spreading fake news in a social media environment to gain likes or followers while maintaining credibility. In the fact-checking role, games like MathE \cite{katsaounidou2019mathe} and Escape the Fake \cite{paraschivoiu2021escape} have players work to identify fake news using verification tools such as reverse image search. 

One example of such single-player games is Bad News \cite{roozenbeek2019fake}. Here, players actively learn the strategies used to create and spread fake news within the game's narrative, such as the use of emotionally charged content and the manipulation of social media platforms. Through these mechanisms, players became aware of the psychological techniques behind misinformation, improving their ability to critically evaluate real-world information. By contrast, in the game Fakey, the goal of the player is to support a healthy social media experience by promoting information from reliable rather than low-credibility sources \cite{micallef2021fakey}. Similarly, in the game Cranky Uncle, players were shown popular misconceptions related to the vaccine and shown how to counter them\cite{cook2023cranky}.

There are also work adapting PvP and team mechanics in serious games about misinformation. FakeYou! is a mobile game where players create fake news headlines and test their ability to spread misinformation by challenging another player's ability to recognize misinformation \cite{clever2020fakeyou}. DoomScroll proposes a team mode where players tackle misinformation challenges together \cite{wells2024sus}.

%ray next
\subsection{AI-Driven Interactions for Misinformation Education}
AI-based technologies currently employed in combating misinformation include automated fact-checking \cite{choi2024fact},
AI-based credibility indicators \cite{lu2022effects}, AI and LLM-based explanations of content veracity \cite{horne2019rating}, and personalized AI fact-checking systems \cite{jahanbakhsh2023exploring}. These efforts focus on debunking interventions, where false information is identified and corrected after dissemination. There are fewer works in prebunking forms of media education, although one study  empowers players to critically engage with misinformation through investigative role-play \cite{tang2024mystery}.

Recent advances in Generative AI (GenAI) LLM-based agents can simulate human behavior based on past events and reflection \cite{zhou_retrochat_2025,he_i_2025,liu_salt_2025}. These agents can be designed for gaming contexts using natural text-based descriptions \cite{ling_sketchar_2024,ma_follow_2025} to support narrative design \cite{fu_being_2024}. Applications of these types of GenAI enabled designs include a text-adventure game where players can freely interact with NPCs generated by GPT-4, leading to emergent gameplay behaviors\cite{peng2024player}, and a GPT dialogue-based game in a speculative post-climate world\cite{zhou2024eternagram}.

However, working with LLMs can lead to ``hallucinations'' when LLM spontaneously produce false information \cite{park2023generative}, as well as bias and stereotype. Attempts to address this include excessively long prompts that come with its own risks of inconsistent outputs \cite{zhong2024memorybank}. Even variations in writing style and spelling in the input text can impact the outputs, leading to generation of incoherent outputs \cite{han_when_2024,chen_once_2025}. These challenges of working with LLMs must be addressed during careful prompting in the game design process. 

\subsection{Mapping Game Design to Media Literacy Constructs} 
Prior misinformation game interventions typically adopt single-player, choice-based formats that aim to affect player recognition abilities rather than production competencies. To address this, we aim to design a game to foster both critical consuming and critical prosuming skills through competitive play. We employed a PvP format with free-form responses, enabling players to craft persuasive or corrective arguments based on both game background news and their opponent’s messages. We integrated LLM-simulated evaluators that dynamically assess player messages, thereby improving critical consuming skills through active interpretation of personalized feedback. Together, these design choices aim to support participants' development of media literacy skills while offering an immersive and replayable learning experience.

\section{Game Design Approach}\label{sec:Game design}
\subsection{Overview of Game Design}
The game we designed focused on the challenges of managing information in a health crisis. One player assumes the role of an Influencer hired by a company to promote a remedy based on traditional medicine but lacking extensive scientific support. This player can create and disseminate misinformation about the remedy. The second player takes on the role of a Journalist advocating for a newly developed medicine supported by scientific research. The goal of this player is to debunk/disprove the misinformation spread by the Influencer. The game features a system that simulates public opinion, whereby an LLM models the reactions of five characters who read players' messages. The objective is to sway the simulated public opinion in favor of one's position.
%Previous studies showed the robustness of the LLM in evaluating the quality of an argument \cite{mirzakhmedova2024large}.

\subsection{Gameplay}
\subsubsection{Game Flow}
\begin{figure}[h]
    \centering
    \includegraphics[width=0.9\linewidth]{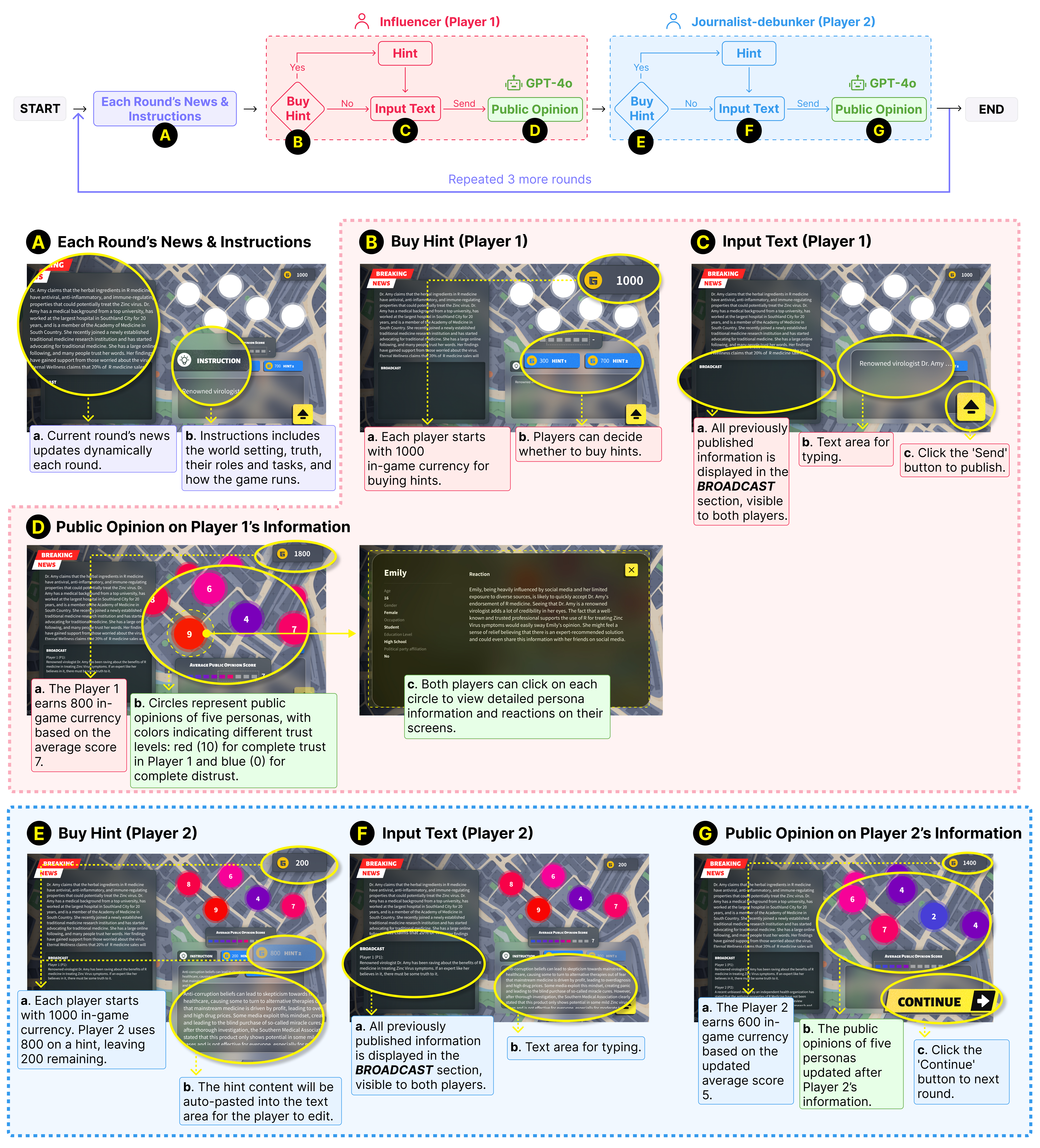}
    \caption{Game Flow. (A) Both players read the current round's news and instructions. (B) Influencer starts first to generate misinformation by choosing whether to buy the hints. (C) Influencer inputs text and sends a request to the GPT-4o API. (D) GPT-4o API returns a response of public opinion. (E) Journalist then starts to counter Influencer's misinformation by choosing whether to buy the hints. (F) Journalist inputs text and sends a request to the GPT-4o API. (G) GPT-4o API returns an updated public opinion.}
    \label{fig:flow}
\end{figure}

Participants were randomly assigned to one of two roles: Influencer or Journalist. Both players are provided with instructions, including the setting, the fundamental reality regarding the effectiveness of the two medicines, their roles and tasks, demographic information about the characters who represent public opinion, and the rules of the game (\autoref{fig:flow}A).

The game unfolds over four rounds, each featuring a new set of updated news. In each round, the Influencer begins by reviewing the news and any instructions. They can also decide to buy hints provided in the game using in-game currency (\autoref{fig:flow}B). Once ready, the Influencer types their information and it is published (\autoref{fig:flow}C). The LLM reacts to this information by simulating public opinion, with this impact of the information on public opinion visible to both players (\autoref{fig:flow}D).

Next, the Journalist takes their turn. Journalist reviews the current public opinion and where appropriate counters any misinformation by typing their debunking response, which they then publish (\autoref{fig:flow}F). If needed, they can purchase customized hints (\autoref{fig:flow}E) or read the instructions. After publishing their response and receiving feedback from the LLM, the round ends, and the game progresses to the next round (\autoref{fig:flow}G).

At the end of each round, both players can view the results, which reflect the Journalist's impact on misinformation. The process for the remaining three rounds is the same as in the first round.

\vspace{2mm}

\subsubsection{Narratives}
The game is set in a fictional small country with called Southland. Historically, Southland has been known for producing renowned medical doctors and pharmacists. However, there are ongoing debates in this country about the comparative merits of modern healthcare methods and traditional medicine. %due to the lack of data on the safety, efficacy, and quality of most medicinal plants. 
The sudden outbreak of the ``Zinc Virus'' further amplifies these debates. As the healthcare system becomes overwhelmed and the scientific community unable to provide an effective treatment because of limited knowledge about this novel virus, residents turn to traditional medicine in search of hope.

We set the game in a health crisis scenario because, in real life, situations marked by scientific uncertainty, where authorities can be unable to provide confident full explanations or advice – often  fuels rumors and speculation about treatments\cite{wang2019systematic}. These dynamics were observed during the Ebola\cite{fung2016social}, Zika\cite{wood2018propagating}, and COVID-19 pandemics\cite{suarez2021prevalence}. In such scenarios, traditional medicine frequently promoted to prevent or treat viruses\cite{lam2021public,mutombo2023experiences}. Additionally, we incorporated the traditional medicine controversy into our narrative as these debates are well-known to our participants, who were of an East Asian background. Research showed that 48.4\% of Hong Kong residents reported using traditional medicine before the COVID-19 pandemic\cite{lam2021public}. Similarly, a national survey in South Korea found a 74.8\% prevalence of traditional medicine use overall\cite{ock2009use}. In China, traditional medicine is formally integrated into the healthcare system as a widely practised modality\cite{chung2023implementation}. 

We use “Product R” to represent traditional medicine and “Max” for science-based medicine. Before gameplay, we explained to players the nature of Product R, particularly stating that it has not undergone rigorous testing and lacks scientific consensus on its safety. Players are also informed that Max has been subject to rigorous clinical trials, with the results published in a peer-reviewed medical journal, however, these studies have demonstrated inconsistent effectiveness against the Zinc virus.
We deliberately avoided making vaccines a topic of this game to prevent players' pre-existing attitudes toward vaccines from influencing their behaviour in the game.

To create the ``News'' pieces in the game, we first researched examples of misinformation by reviewing relevant literature. We selected key features of misinformation and incorporated them into the game's events (See details in \ref{Characteristics of Misinformation}). To ensure the misinformation was portrayed realistically, we investigated real-world examples from fact-checking websites\footnote{\url{https://www.snopes.com/}}, reputable news outlets\footnote{\url{https://www.wsj.com/}}, and medical websites\footnote{\url{https://www.webmd.com/}}\footnote{\url{https://sciencebasedmedicine.org/}}\footnote{\url{https://healthfeedback.org/}}. For instance, we represented ``less credible sources'' using personal stories, viral videos, and newspapers that can objectively be classified as being biased. The portrayal of biased newspapers draws on research showing that political orientation can influence medical preferences. For example, research has shown that voters who tend to support anti-corruption parties are less likely to seek services from mainstream healthcare providers, and are more inclined to use alternative treatments\cite{valtonen2023political}.  
Next, we crafted ``News'' pieces for the game based on these findings. However, any similarities to real-world sites or companies are purely coincidental; all names and events was invented solely for the purpose of this study. \autoref{Misinformation-features-Table} summarizes the key features of misinformation identified in the literature, the associated cited papers, and how these are reflected in the in-game news. The full version of the ``News'' is in the supplementary materials. Lastly, To ensure balanced gameplay, we used a GPT-4o model to review the narrative and provide an opinion on the difficulty to players of dealing with each piece of news in the game context. Taking this opinion into account, we made further revisions and corrections when we conducted two pilot tests with four people. The goal was to ensure the game was balanced and gave opportunities to win the game for both players.

\begin{table*}[t]
\renewcommand\arraystretch{1.35}
\vspace{2mm}
\centering
\caption{Key misinformation features and corresponding ``News'' in the Game}
\begin{tabular*}{\textwidth}{@{}P{5cm} P{8cm}@{}}
\toprule
\textbf{Misinformation Characteristics} & \textbf{Representation in the Game's News} \\ 
\midrule
False information is often shared by lower-quality media. Political ideology, however, shapes people's perceptions of trustworthiness.\cite{zhang2020overview,molina2021fake,hanley2023golden}
& 
A newspaper reported that a renowned medical expert has advocated Product R, claiming its herbal ingredients could potentially treat the Zinc Virus. This newspaper is known for its anti-corruption stance. (Round 1) \\
\midrule
Personal, negative, and opinionated tones predominate in misinformation narratives which frequently provoke dread, anxiety, and mistrust of institutions.\cite{bessi2015trend,porat2019content} 
& 
A widow shared her husband's experience. She suspects that Max was ineffective and believes it may have caused renal impairment, eventually leading to her husband’s death. She claims, “He was given a medication we demanded he NOT receive, and his health quickly went downhill,” ultimately resulting in him being “on a ventilator working most of the time at 100\%.” (Round 2) \\
\midrule
Fake news, some of which is purposely fabricated to cause harm, generate financial returns, or spread confusion.\cite{mustafaraj2017fake,hanley2023golden,shu2017fake} 
& 
A Journalist discovered that the institution of traditional medicine where the famous medical expert works received significant funding from billionaire Jack. Additionally, Jack's ex-wife owns a company that produces and promotes traditional medicine products like R. (Round 3) \\
\midrule
False rumors will create feedback loops and evolve into more intense and extreme versions over time.\cite{shin2018diffusion} 
& 
A popular short video claims that a doctor who practiced alternative medicine and R was murdered to protect the profits of “Big Pharma”. More people are attracted to believe in the validity of traditional medicines and advocate for their use while opposing new drugs. Growth in sentiment that resistance to traditional medicine amounts to being an attack on their cultural heritage. (Round 4) This round's advocate for R becomes more intense than in Round 1, with the focus being less on effectiveness but rather patriotic sentiment.\\
\bottomrule
\label{Misinformation-features-Table}
\end{tabular*}
\end{table*}

\subsubsection{Instructions and Hints}
To support players, the game includes instructional content that features definitions\cite{molina2021fake,wu2019misinformation}, examples\cite{bbcCoronavirusBill}, and strategies for both creating and debunking misinformation. This content draws on insight gained into misinformation from research literature and practice. For the Influencer, we applied the Elaboration Likelihood Model and used simple examples to teach players to craft persuasive misinformation\cite{petty1984source,moran2016makes}. For the Journalist, we used an Agence France-Presse fact-checking style-guide and an guide published by the EU on communicating with proponents of conspiracy theories \cite{afp2024,eeas2024}. These were the inspiration for a user-friendly guide we developed to assist game players to identify misinformation and equip them with effective debunking strategies. 
Additionally, each round offers two hints to support players. The detailed hint is crafted by the authors using the same materials as the ``News''. The simple hint is generated by GPT-4o model. When authors used it to review the news and ensure balanced gameplay, it provided concise suggestions on how each player might respond from their perspective. (Full instructions and hints can be found in the supplementary materials)

\subsubsection{LLM Basis}
We implemented the LLMs to play the role of ``public opinion'' in the game for three reasons.  Firstly, LLMs perform well when processing dynamic natural language\cite{hu2024survey}. Secondly, LLMs demonstrate memory capability, such as with working memory being applied to the context of a conversation, and long-term memory allowing past conversational information to be taken into account\cite{hu2024survey}. 
 Research also demonstrates that generative agents powered by LLMs build a high degree of capability for responding to the context of a conversation\cite{park2023generative,yin2024lies}. Thirdly, LLMs excel at role-playing tasks in the game. Research indicated\cite{wang2023humanoid}\cite{hu2024survey} that directly inserting natural language descriptions of a role’s identity enable LLMs to make better quality evaluations in conversational tasks. These capabilities allow LLMs to effectively serve as ``evaluators'' in the game, generating continuous, context-aware dialogue and feedback. This approach proves more effective than traditional prebuilt game mechanics, such as trigger keywords for assessment.

\subsubsection{Game Mechanics and Interface}
\begin{figure}[htbp]
    \centering
    \includegraphics[width=1\linewidth]{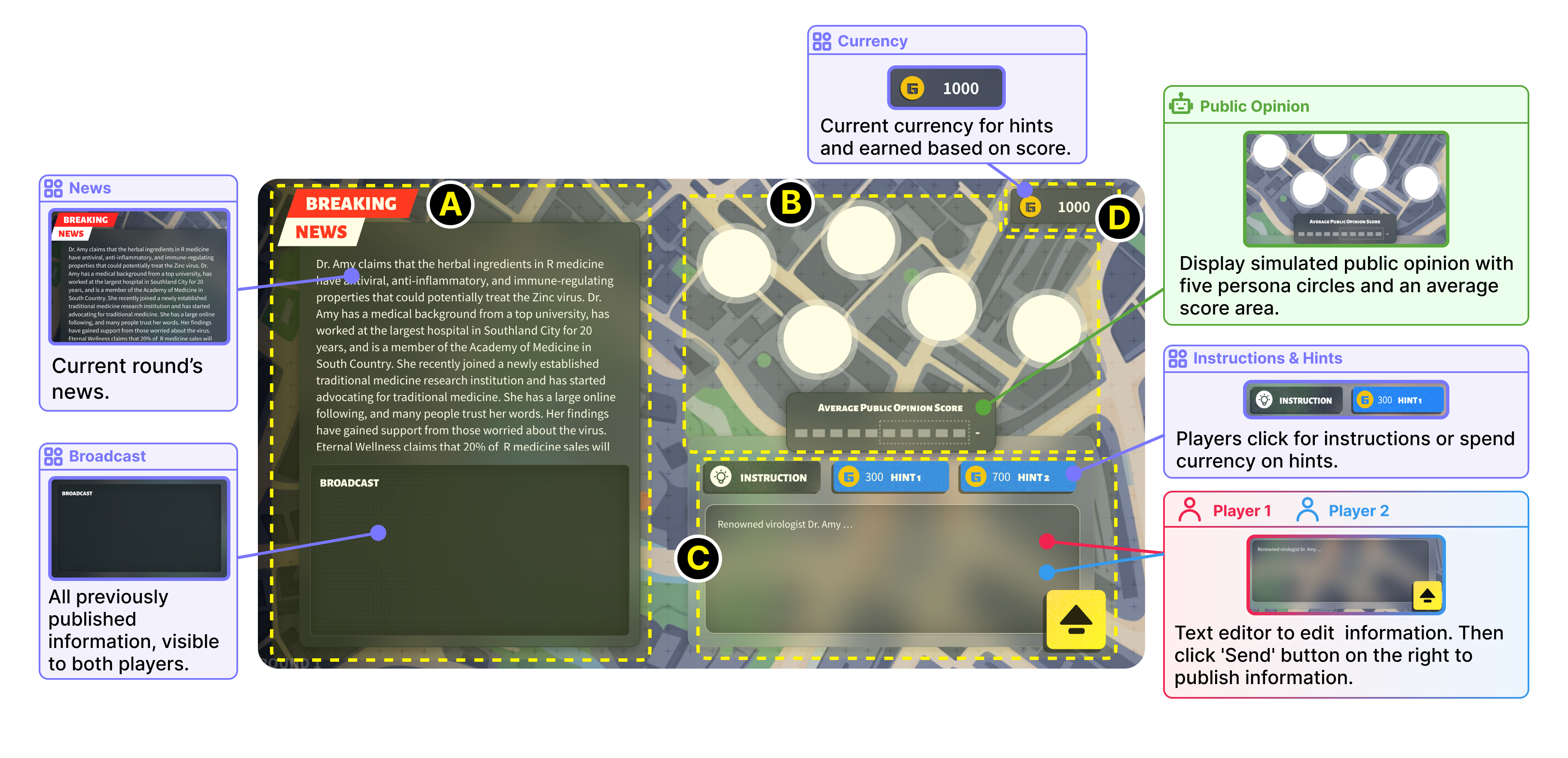}
    \caption{Game Interface. (A) Players can view the current round's news and all previous information published by both players. (B) Players can view LLM-simulated public opinion information. (C) Players can edit their information in the text editing area, view instructions and buy hints. (D) Players can view their own holdings of in-game currency.}
    \label{fig:interface}
\end{figure}

%Building on this LLM cability, we designed the game mechanics to foster player understanding of misinformation generation and debunking through interactive competition. Players take on distinct roles: the Influencer produces misinformation, while the Journalist counters with corrections. Each round, players submit messages, which are evaluated by five AI personas acting as the simulated public. These personas respond with trust level scores reflecting message credibility, and the average trust level score determines round outcomes. The player's success depends on the reaction of LLM-simulated public opinion. 
%The scoring system helps players adjust their strategies for the next rounds. To keep players engaged, they earn in-game currency each round based on their performance reflected by average score, and can also use it to purchase hints. Full details of the scoring system and in-game currency rules are provided in the Appendix.
The text-based game interface having four sections that allows players to view the information clearly for decision-making during gameplay. Players can view the current round's news and all previously published information in the Information Viewing Section (\autoref{fig:interface}A); they can also see the LLM-simulated public opinion in the Public Opinion Section (\autoref{fig:interface}B). In the Text Editing Section (\autoref{fig:interface}C), players can edit their information, view instructions, and purchase hints. Player success is determined by the reaction of LLM-simulated public opinion, measured through trust level scores. This scoring system helps players adjust their strategies in subsequent rounds. To maintain engagement, the game features a reward system tied to trust scores: players earn in-game currency each round based on their average score, which can be used to purchase hints. Full details of the scoring and currency system are provided in the Appendix.

\begin{figure}[htbp]
    \centering
    \includegraphics[width=1\linewidth]{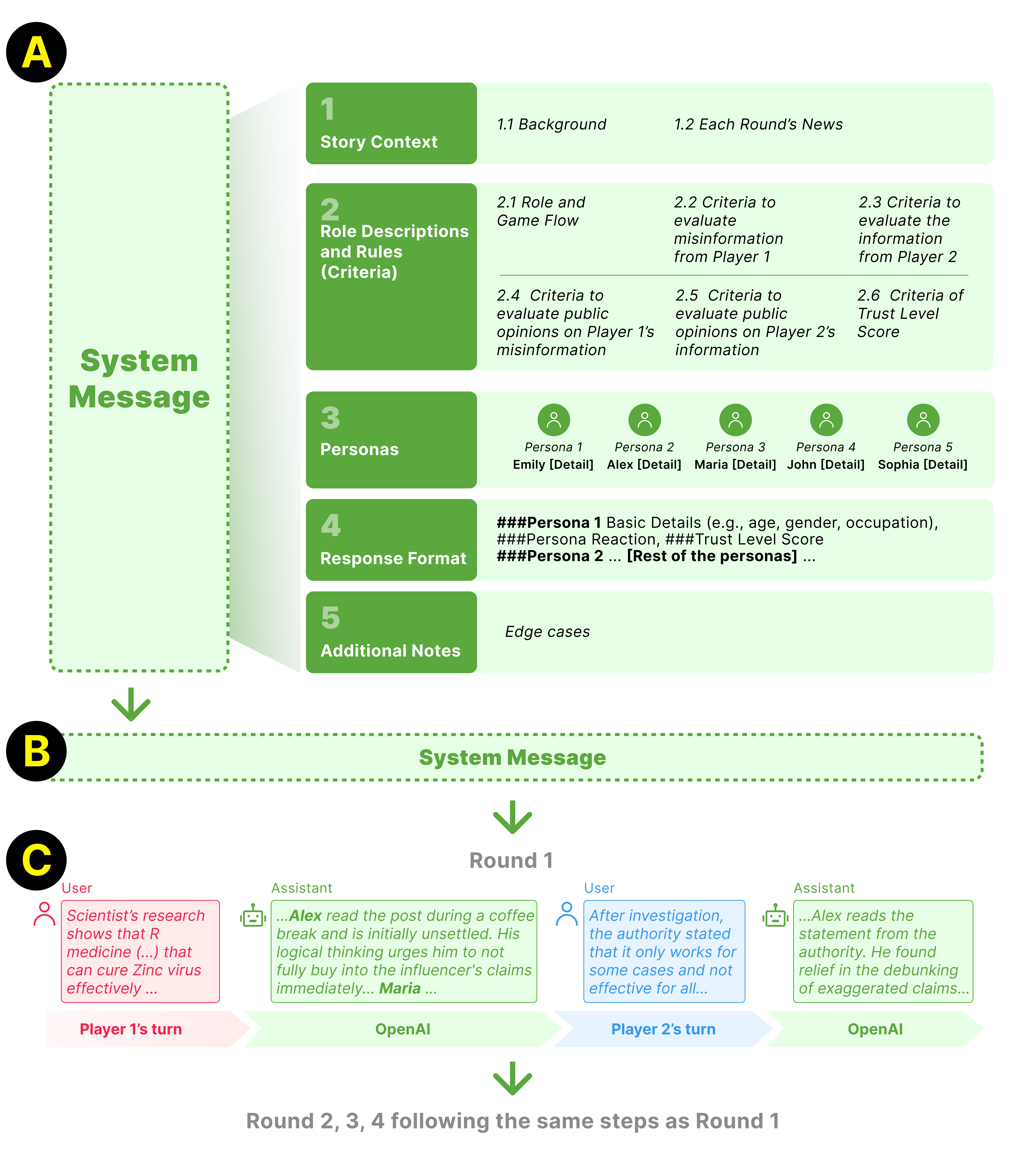}
    \caption{Prompt and Workflow.(A) The prompt in the game is structured into five sections: Story Context, Role Descriptions and Rules (Criteria), Personas, Response Format and Additional Notes. These five sections form the System message (B). Once the system message is completed, it is applied into the game (C). During gameplay, the LLM generates and simulates public opinions based on the previously established system message. After four rounds, the game ends.}
    \label{fig:prompt}
\end{figure}
\subsection{Prompt Engineering }
\label{Prompt Engineering}
%The configuration of prompt engineering is to prompt the GPT-4o model, helping it understand its tasks based on our game’s context, mechanics and flow. This is critical to more accurately simulate public opinion based on the persona provided, including trust level scores and reactions, upon receiving player input. 

We employed the GPT-4o model and integrated prompt design techniques informed by prior work. The prompt was structured into four sections:
\begin{itemize}
    \item First, we provided a game story context (\autoref{fig:prompt}-A-1) to establish a consistent narrative foundation. This ensured that all outputs generated by the LLM remained coherent and aligned with the game’s world setting.
    \item Second, we assigned the LLM a specific role (\autoref{fig:prompt}-A-2), clearly defining its responsibilities and tasks. As the core section of the prompt, we included detailed instructions for the LLM. To guide the model effectively, we applied the Rails approach\cite{amatriain2024prompt} by predefining rules to constrain the LLM’s output. Additionally, we used the Chains approach\cite{amatriain2024prompt} to structure the workflow clearly to let the LLM process the task step-by-step.
    \item Third, since the core game mechanic involves the LLM generating diverse public opinions, we used the Expert prompting approach\cite{amatriain2024prompt} and designed five distinct fictional personas (\autoref{fig:prompt}-A-3). This helped the LLM create content that matches each persona’s perspective. We further enriched the personas by incorporating insights from the literature and focused on four key group factors: demographics, psychological traits, personality, and behavioral features. These factors guided how each persona responded to misinformation and anti-misinformation messages.\cite{nan2022people,liu2023checking,lee2024misinformation,shin2024understanding}. 
    \item In section 4, we reinforced output consistency and quality by including explicit output format rules and example prompts (\autoref{fig:prompt}-A-4). These examples taught the LLM\cite{amatriain2024prompt} how to produce coherent, high-quality outputs aligned with the game design.
\end{itemize}
A more detailed version of the prompt structure can be found in the \hyperref[Prompt Design]{appendix} .

\subsection{Implementation}
\subsubsection{Multiplayer Setup}
\begin{figure}
    \centering
    \includegraphics[width=1\linewidth]{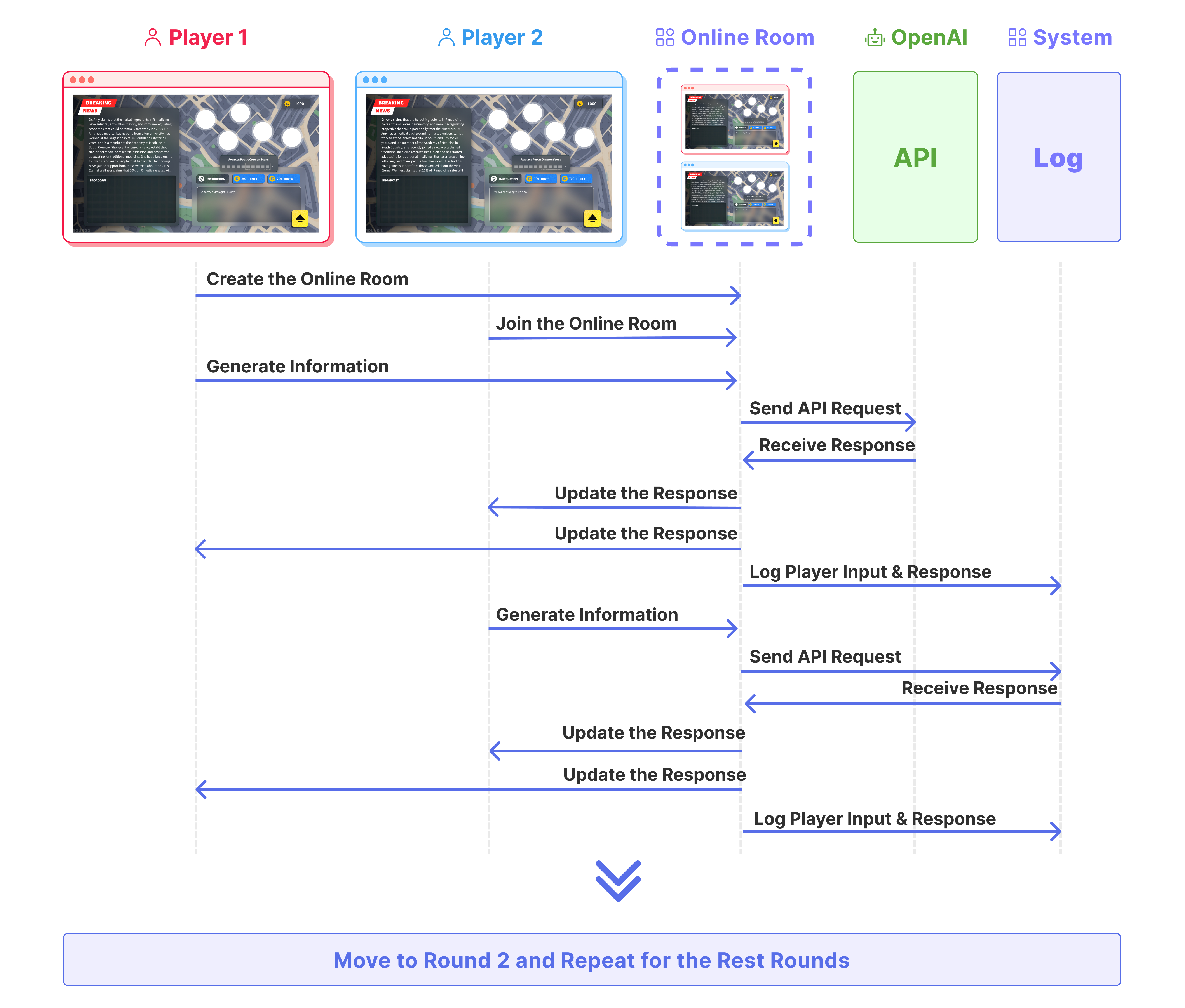}
    \caption{Interaction System. Player 1 creates the online game room as the host and Player 2 joins. After Player 1 sends a request to GPT-4o API, Player 2 can view it. Both players in the online room receive the API response, which is processed and updated on their own game screens. Then Player 2 takes their turn to input information, following the same process as Player 1. At the end of the round, all in-game events (including player inputs, API responses, time, etc.) are logged locally.}
    \label{fig:system}
\end{figure}

To support the multiplayer functionality, we used Photon Unity Networking (PUN). Photon enables real-time multiplayer interactions by providing the network server connections to players, thus creating a shared game state that is synchronized across all clients. The game begins by establishing an exclusive online Photon room (\autoref{fig:system}), where only participants can join and interact. % Photon’s cloud-based architecture ensures low latency, reliable connections, %even when players are geographically away .
In this configuration, critical game variables and data (such as the player’s actions, messages, and game state) are synchronized across both players' screens using Photon’s Remote Procedure Calls (RPCs). This synchronization ensures that any action taken by one player is immediately reflected and displayed on the other player’s screen.%, such as updating the game’s UI elements like input fields, buttons.

\subsubsection{Data Storage}
The game employs a logging system to store and manage game data locally on the player's device (\autoref{fig:system}). The log files record various in-game events, including player inputs, API responses, and game state changes. This data is used for analyzing players’ behavior, such as how players interact with the game. To make the data easy to access and ensure compatibility on cross-platform such as Windows and macOS, the log files are saved in a created folder named GameLogs under the players’ commonly used directory. This directory structure is automatically created when the game starts.% ensuring that all log files are organized and easily retrievable.

\section{Methods}\label{sec:Methods}
To address our research questions, we used a mixed-methods within-subjects study design.
\subsection{Evaluation Methods}

\subsubsection{Key Concepts and Measurements}
This paper focuses on the effectiveness of the game as a prebunking intervention in enhancing participants' skills to protect themselves against misinformation. Following Lewandowsky and Van der Linden \cite{lewandowsky2021countering}, we conceptualise \textbf{prebunking interventions} as follows: interventions which contribute to a person's resilience to misinformation via raising critical media literacy skills (media knowledge and critical assessment of information), resulting in the will and ability to apply this knowledge to make decisions about information veracity. In contrast, debunking interventions can be understood as any intervention, which aims to correct the person's opinion about information post hoc (after the person was exposed to and believed in) specific misinformation. %; it is also implicitly listed that while the debunking interventions, which incorporated reasoning elements in generally works better, the main goal of debunking intervention is to inform, rather to teach person.} 

In this paper, we distinguish between the concept of ``debunking'' (the practice of challenging or disproving misinformation), and ``debunking interventions'' (social media platforms' actions taken to perform debunking for user). We do not discuss ``debunking interventions'' (e.g., content labelling \cite{lewandowsky2021countering} or Community Notes \cite{chuai2024community}); instead, \textbf{we focus on ways to improve individuals' debunking skills as a result of prebunking interventions}.

Based on the definitions of prebunking interventions, it has become clear that we need to assess the educational effect of the intervention in relation to a) media literacy skills/knowledge gained in the intervention, b) behavioral intentions to apply that knowledge, and c)  the practical ability to use this knowledge to recognize misinformation. Together, it would help to understand the effectiveness of the intervention and estimate the increase in debunking skills of the participants. This approach is similar to Tifferet \cite{tifferet2021verifying} taxonomy for measuring susceptibility to misinformation, which can be viewed as the inverse of the resilience fostered by prebunking. 

%Based on the definitions of prebunking interventions, it is become clear that %while efficiency of debunking interventions can be measured through opinion change after being exposures to the debunking materials (e.g. as it was performed by many papers on the subject discussed in\cite{martel2023misinformation}), 
%we need to assess educational effect of the intervention in the relation to a) how much the media literacy skills/knowledge are progressed after the intervention, b) how much person would like to implement this new knowledge in practice c) and how this knowledge is working to help people recognise the examples of misinformation in practice to measure. Together, it would help to understand the effectiveness of the intervention and estimate the increase in debunking skills of the participants. This approach is similar to Tifferet \cite{tifferet2021verifying} taxonomy of the evaluation of susceptibility to misinformation, which can be viewed as the inverse of the resilience fostered by prebunking. 

Tifferet categorizes measurements into three main groups: performance tasks (how well users can discriminate between fake and real pieces of news), general media literacy assessment (how much a person knows about different aspects of misinformation), and behavioral assessment (how much a person would like to use different strategies to verify online information) \cite{tifferet2021verifying}. As Tifferet argues, these three aspects are complementary to understanding user’s susceptibility to misinformation and, therefore, evaluation of improvement in each of them could provide a full picture of our game's efficiency.
In addition, as the game was designed to present complex scenarios, we decided to evaluate how confident participants were in their ability to recognise misinformation both before and after the intervention.

In the study, we used the validated questionnaires dedicated to measuring each of these aspects.

\subsubsection{Questionnaires}
\textbf{Media Literacy Assessment: New Media Literacy Scale (NMLS)}.
To assess changes in media literacy, we used the NMLS \cite{koc2016development}. The scale is designed to measure literacy in ``new media'' (digital media and social networks) and is based on the four factors from the model by Chen et al.: functional consuming, critical consuming, functional prosuming, and critical prosuming \cite{chenwu2011unpacking}. 

The NMLS was selected primarily because it assesses the critical dimensions of content production and consumption, which are core competencies our intervention aimed to establish. Notably, compared to other media literacy instruments \cite{eristi2017development, vraga2015multi, ashley2013developing}, the NMLS is the only scale which provides this dimension. Additionally, the scale was developed and validated on a group of university students, which reflected our projected sample. The questionnaire includes 35 questions, rated on a 5-point Likert scale. We expected our intervention to enhance critical prosuming skills (as players were required to create influential content and sway LLM opinions) and critical consuming skills (as players had to analyze information from the game and other players to craft effective responses). We also anticipated positive effects on lower-level functional production and consumption skills (basic understanding of media consumption and creation), as the game offered basic training in reading, understanding, and responding comprehensively to media texts.

\textbf{Behavioural Assessment: Verifying Online Information Scale (VOI - 7)}.
To assess the effect of the game on the verification practices performed by the participants, we adapted the VOI proposed by Tifferet \cite{tifferet2021verifying}. 
To the best of our knowledge, this is the only existing scale which focuses on the behaviors (verification practices) a person can adopt to verify the news.
The questionnaire measures individuals' differences in applying direct and indirect verification practices for online information, allowing us to track expected behavior changes in verification practices. We used the VOI-7 version, which demonstrated comparable construct characteristics to the original 22-question version while allowing to be completed more rapidly. The parameters were measured on a slider from 0 to 100, where participants were asked to indicate their likelihood of applying verification practices.
As our game learning materials and the gaming procedure show the importance of verifying the information (via showing multiple misinformation-related events and presenting features of misinformation which should be checked to avoid being misled, we expected, that people will be more willing to apply verification practices.

\textbf{Performance Assessment: Misinformation Susceptibility Test (MIST - 20)}. To assess changes in veracity discernment, we used the MIST \cite{maertens2024misinformation}. 
To date, the MIST is the only fully validated misinformation susceptibility instrument. It takes into account the ability to recognize real news and fake news presented in equal proportion. The MIST framework is designed to allow for the comparison of results across different studies and interventions. The test has been implemented in multiple misinformation intervention assessment studies (e.g. \cite{roozenbeek2022psychological,spampatti2024psychological}), including media literacy/misinformation games  \cite{bradstreet2023data,wells2024doomscroll} and having considerable predictive validity \cite{maertens2024misinformation}, therefore giving comprehensive estimates of people's ability to recognize real misinformation. We applied the MIST-20 version, which includes 20 items. Participants were asked to rate each item as either a ``fake'' or ``real'' news headline.
We anticipated that participants would learn heuristics for identifying potential misinformation through intense interaction with the game. This interaction, which included hints about the characteristics of misinformation and the cognitive work of creating or debunking it, was expected to facilitate this learning.

%\textcolor{blue}{As we expected, because of intense interaction with game news, which contained hints to how misinformation works, and because of the cognitive work of creating or debunking misinformation, participants learned the heuristics to identify potential misinformation content.}

\textbf{{Self-efficacy Assesement: }Fake News Self-efficacy Scale}. 
To measure perceived self-efficacy in dealing with fake news, we used a 3-item questionnaire \cite{hopp2022fake}. This questionnaire assessed participants' confidence in three key areas: (1) their ability to identify news-like information that may be intentionally misleading, (2) their ability to distinguish between fake news and content produced with honest intentions, and (3) their ability to recognize news that may be unintentionally incorrect (i.e., misinformation). We chose the scale as a better alternative to the non-validated single-item measurement of confidence in identifying fake news, used by \cite{hinsley2021fake}.
Each item was rated on a seven-point scale. 

\subsubsection{Qualitative Data}
%\textcolor{red}{\sout{\subsubsection{Semi-structured Interview}}

\textbf{Semi-structured Interview: }To evaluate the user's experience in-depth, connected with the content of the game and the strategies implemented by users, we developed a protocol for a semi-structured interview. This protocol includes questions about the general experience, the perceived goal of the game, the perception of the opponent's strategies, and the individual's perception of the game's effectiveness or ineffectiveness. The guidelines for the semi-structured interview are presented in the Supplementary material.

%\textcolor{red}{\subsubsection{Log analysis}}
\textbf{Game Log: }The game logs collected the data including player-generated content, the time spent in each round, API responses showing the public opinion of different personas, trust level scores, and in-game events such as the amount of money players had, how much they spent, and what hints they purchased. This data provided a %precise 
transcript of each session, enabling the research team to analyze players’ strategies, in-game behaviors, and decision-making processes.

\textbf{Quantitative Data Analysis:} We employed a combined inductive-deductive approach to analyze the interview transcripts and gameplay logs\cite{kuckartz2019analyzing}. This approach ensured a comprehensive understanding of the gameplay experience. Our primary objectives were to gain understanding of how participants perceived and understood misinformation through the game, how they learned to distinguish and apply debunking strategies during gameplay, and how interactions with other players influenced their behavior and learning. The analysis process began with inductive coding. Two researchers independently coded a subset of the data, identified themes, and then discussed and reconciled any coding discrepancies, iterating on the coding system as needed. Once the coding system was established, the two researchers independently coded the full dataset. A third researcher then reviewed the coded data, and any differences in interpretation were discussed until a consensus was reached.

\subsection{Recruitment and Participants}
Participants were recruited through flyers and university-affiliated online media groups. We also encouraged participants to share information about the study within their networks. The eligibility criteria required participants to be adults and have sufficient English proficiency to play the game (we also do not forbid using translation engines if any of the aspects of the game are not understandable). Given that the proliferation of online misinformation is a global challenge and commonly reaches unsuspecting users\cite{ferrara2020misinformation}. % even if people encounter misinformation, they do not necessarily know about it if the information is debunked later. Thus 
We did not require participants to have prior exposure.
%Instead, we assumed they had already encountered misinformation, especially in the aftermath of the COVID-19 pandemic.

60 participants initially expressed interest in participating in the game intervention study. Ultimately, 47 participants were selected, forming 24 pairs for the game sessions. In one of the pairs, one of the study's authors participated in a player role due to scheduling reasons. Because the data collection form allowed participants to skip questions, 3 participants did not complete the entire MIST questionnaire, and 5 participants left some questions blank in the pre-procedural VOI questionnaire. Their data were excluded from the VOI and MIST data analyses. For the control condition, 57 participants signed up, and 50 successfully completed both the pre and post questionnaire and the reading exercise.

Demographic characteristics of the participants are summarized below. In the game intervention group, the participants’ ages ranged from 20 to 57, with a mean age of 25.87 years (SD = 6.265). 28 participants identified as female, 18 as male, and 1 preferred not to disclose their gender. 3 participants reported having an Associate degree, 29 a Bachelor's degree, and 15 a Master's degree; all participants reported having Eastern Asian origin. Participants in the control group ranged in age from 21 to 42, with a mean age of 27.46 (SD = 6.45). Thirty-three participants self-identified as female and 17 as male. Seventeen held a Master’s degree, 27 a Bachelor’s degree, and 6 an Associate degree. All participants reported being of East Asian origin. (See participant demographic information in Appendix \ref{Demographic Information of Participants})

%The study sample had the following characteristics: the participants’ ages ranged from 20 to 57, with a mean age of 25.87 years (SD = 6.265). 28 participants identified as female, 18 as male, and 1 preferred not to disclose their gender. 3 participants reported having an Associate degree, 29 a Bachelor's degree, and 15 a Master's degree; all participants reported having Eastern Asian origin (see participant demographic information in \ref{Demographic Information of Participants}).

The experimental design was approved by the Ethics Review Panel of City University of Hong Kong. As the game story was centered around a fictional pandemic, we informed participants about the theme in the consent form and asked them not to participate in the study if they perceived the topic of health/diseases to be disturbing.  All participants gave their informed consent and were compensated 40 Chinese Renminbi upon completion.

\subsection{Procedure}
\begin{figure}[htbp]
    \centering
    \includegraphics[width=1\linewidth]{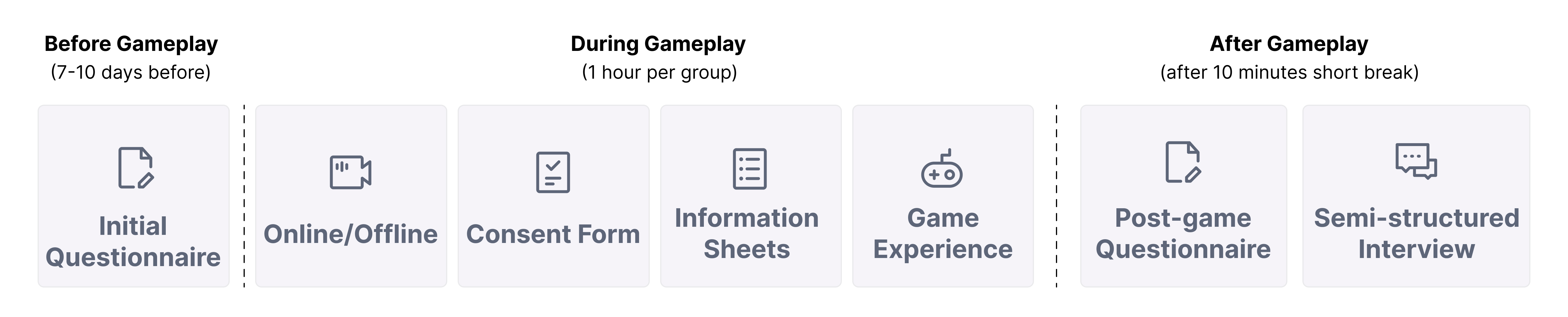}
    \caption{Overview of Game intervention study procedure.}
    \label{fig:study procedure}
\end{figure}

%with repeated measures.
Once a person expressed interest in participating, they completed an initial questionnaire containing demographic questions and measurement scales. To prevent participants from intentionally biasing their responses, the questionnaire was administered 7-10 days before the gameplay experiment. After participants confirmed completion of the questionnaire, we scheduled the gameplay sessions. These experimental sessions were conducted either online via the VooV Meeting application or in person at a university meeting room. At the start of their test session, participants were given information sheets and consent forms to review and complete at their own pace. Once completed, Participants were introduced to the game setup and roles, and when they decided between themselves which role they would like to play. After the gaming session, participants again filled out the questionnaires. Finally, we conducted a short semi-structured interview to discuss their perceptions of the game. The entire session last approximately one and a half hours (See \autoref{fig:study procedure}). 
%including an additional questionnaire about their game experience

\subsection{Control Group}
To verify whether the post-game improvements in the questionnaire results were not simply due to the Hawthorne effect \cite{mccambridge2014systematic}, and to test whether passive exposure to misinformation alone could improve understanding, we conducted a control group study. %and to demonstrate that mere exposure to misinformation is not sufficient to improve understanding, we implemented a control group. 
Participants in the control group were asked to read examples of misinformation-containing social media content. The reading materials were selected from the Fake News Dataset\cite{DBLP:conf/clef/KohlerSSWS0S22}, which contains fact-checked claims and URLs from various fact-checking websites. We chose 40 posts related to the same topics used in the game intervention condition(themes regarding health, personal experience, alternative medicine, and pandemic). The estimated reading time was approximately one hour, which matched the expected duration of the game intervention. Before the task, we explicitly informed participants that the materials presented were examples of misinformation and that the purpose of reading them was to better understand how misinformation operates. Participants completed the same pre- and post-surveys used in the intervention group. The pre-survey was administered 7–10 days before the reading task. After completing the reading, participants took a short break, then proceeded to complete the post-survey.

%(similar to the ones we used for game preparation by topics and misinformation features in the texts) for a duration equivalent to the expected duration of the game intervention. 
%As in the intervention group, the participants in the control group completed the questionnaires twice: once a week before the intervention and once a week after the intervention. 50 people completed both questionnaires and reading excercise}

%\textcolor{blue}{\subsection{Analysis Protocol}}
%\subsubsection{Quantitative Analysis}

%\subsubsection{Qualitative Data Analysis}

\section{Results}\label{sec:Results}

\subsection{RQ1: Effect on the Ability to Recognise and Understand Misinformation}
\subsubsection {Descriptive Statistics}
%The sample consisted of participants with the following characteristics: the mean age was 25.85 years (SD = 6.265). 28 participants identified as female, 18 as male, and one chose not to disclose their gender. 29 participants held a bachelor's degree, 15 held a master's degree, and 3 held an associate degree. All participants were of Eastern Asian descent. The sample included a diverse range of professional backgrounds, as detailed in the demographic information table.
%The study sample had the following characteristics: the participants’ ages ranged from 20 to 57, with a mean age of 25.87 years (SD = 6.265). 28 participants identified as female, 18 as male, and 1 preferred not to disclose their gender. 3 participants reported having an Associate degree, 29 a Bachelor's degree, and 15 a Master's degree; all participants reported having Eastern Asian origin (see participant demographic information in \ref{Demographic Information of Participants}).

%\textcolor{blue}{The ages of participants in the control group ranged from 21 to 42, with a mean age of 27.46 (SD = 6.45). Thirty-three participants self-identified as female and 17 as male. Seventeen held a Master’s degree, 27 a Bachelor’s degree, and 6 an Associate degree. All participants reported being of East Asian origin.}

Preliminary data analysis reveals significant violations of the normality assumption in pre-tested NMLS scales. Considering the rather small dataset and the Likert-scales-based questionnaires used for most of the scales, we decided to proceed with a non-parametric repeated measures approach (Related-Samples Wilcoxon Signed Rank Test to the results of the scales' pre- and post-evaluations). The scales' descriptive statistics and normality tests results can be found in Table \ref{tab:descriptscales}

% \usepackage{tabularray}

%\begin{longtable}{lll|lllll}
%\caption{Descriptive Statistics and Normality Test Results for Variables Before and After Playing}
%\label{tab:descriptscales} \\

%\toprule
 %& \multicolumn{2}{c|}{\multirow{2}{*}{Mean (SD)}} & \multicolumn{5}{c}{Normality tests (Shapiro-Wilk)} \\
 %& \multicolumn{2}{c|}{} & \multicolumn{2}{c}{W} & \multirow{2}{*}{DF} & Significance &  \\ 
%\cmidrule(lr){4-5}
%\cmidrule(lr){7-8}
%\endfirsthead

%\toprule
% & \multicolumn{2}{c|}{Mean (SD)} & \multicolumn{2}{c}{W} & DF & Significance &  \\
%\cmidrule(lr){2-3} \cmidrule(lr){4-5} \cmidrule(lr){6-8}
%\endhead

%\midrule
%\multicolumn{8}{r}{\textit{Continued on next page}} \\
%\endfoot

%\bottomrule
%\endlastfoot

 %& Before & After & Before & After &  & Before & After \\ \hline
%Functional consuming & 23.09 (0.528) & 24.02 (0.466) & 0.868 & 0.974 & 47 & \textbf{<.001} & 0.378 \\
%Critical consuming & 44.04 (1.056) & 47.74 (0.795) & 0.863 & 0.971 & 47 & \textbf{<.001} & 0.285 \\
%Functional prosuming & 28.02 (0.818) & 28.64 (0.658) & 0.898 & 0.954 & 47 & \textbf{<.001} & 0.060 \\
%Critical prosuming & 36.04 (1.028) & 37.91 (0.920) & 0.941 & 0.958 & 47 & \textbf{0.019} & 0.092 \\
%Self-efficacy & 15.17 (0.445) & 15.26 (0.056) & 0.970 & 0.959 & 47 & 0.265 & 0.101 \\
%VOI & 416.79 (116.459) & 485.93 (128.536) & 0.978 & 0.960 & 42 & 0.588 & 0.153 \\
%MIST & 11.5 (2.162) & 12.5 (2.529) & 0.969 & 0.965 & 44 & 0.284 & 0.208 \\

%\end{longtable}

\begin{table}
\small
\caption{Descriptive Statistics and Normality Test Results for Variables Before and After Playing}
\label{tab:descriptscales}
\centering
\begin{tabular}{lcc|ccccc}
\toprule
 & \multicolumn{2}{c|}{Mean (SD)} & \multicolumn{2}{c}{W} & DF & \multicolumn{2}{c}{Significance} \\
 & Before & After & Before & After &  & Before & After \\
\midrule
Functional consuming & 23.09 (0.528) & 24.02 (0.466) & 0.868 & 0.974 & 47 & \textbf{<.001} & 0.378 \\
Critical consuming & 44.04 (1.056) & 47.74 (0.795) & 0.863 & 0.971 & 47 & \textbf{<.001} & 0.285 \\
Functional prosuming & 28.02 (0.818) & 28.64 (0.658) & 0.898 & 0.954 & 47 & \textbf{<.001} & 0.060 \\
Critical prosuming & 36.04 (1.028) & 37.91 (0.920) & 0.941 & 0.958 & 47 & \textbf{0.019} & 0.092 \\
Self-efficacy & 15.17 (0.445) & 15.26 (0.056) & 0.970 & 0.959 & 47 & 0.265 & 0.101 \\
VOI & 416.79 (116.459) & 485.93 (128.536) & 0.978 & 0.960 & 42 & 0.588 & 0.153 \\
MIST & 11.5 (2.162) & 12.5 (2.529) & 0.969 & 0.965 & 44 & 0.284 & 0.208 \\
\bottomrule
\end{tabular}
\end{table}

\textbf{Effect of the Game on Media Literacy Skills:}
To measure the effects of the game on Media Literacy Skills, we first ran the Related-Samples Wilcoxon Signed Rank Test on the full scale. Then, to determine which components of Media Literacy were most affected by the game, we conducted separate subscale tests to analyze changes in each of the four subdomains of Media Literacy. The results demonstrated significant differences in Media Literacy scale results (N = 47, Z = 3.083, p = .002). The analysis revealed the following differences: the game significantly improved both functional consuming   Z = 2.064, p = .039 and critical consuming Z = 3.344, p <.001 ), but not the functional prosuming Z = .435. p = .664 and critical prosuming Z = 1.868, p = .062. Therefore, the results suggest the game improves Media Literacy in the domains connected to understanding the content of the media and being able to critically evaluate the content of the media; however, it has not significantly improved the ability to produce media content which can be influential to others and support author's ideas \cite{koc2016development}.

In contrast, we did not find the effect of intervention in the control group on any of the parameters media literacy: functional consuming (N = 50, Z = .991, p = .322), critical consuming (N = 50,Z = 1.505, p = .132), functional prosuming (N = 50,Z = 1.474, p = .141) and critical prosuming (N = 50, Z = 1.808, p =.072), showing that merely demonstrate the misinformation content is probably not enough to raise critical approach to media

\textbf{Effect of the Game on the Verification Practices (VOI-7)}:
%To determine if the game practice significantly changed the intention to apply the direct and indirect verification practices, we ran Related-Samples Wilcoxon Signed Rank Test. 
The test revealed significant differences between pre and post-gaming VOI scores (N = 42, Z = 4.361, p < .001). The results suggested that the game positively affected the repertoire of used practices and/or the perceived will to use these practices. 

However, we also found significant differences between pre- and post-scores in the control group intervention (N = 50, Z = 2.208, p = .027), meaning exposure to the misinformation examples also makes people more vigilant and support checking intentions

\textbf{Effect of the Game on Self-efficacy towards Misinformation:}
We did not find significant differences in self-efficacy between pre and post-game measurements (N = 47, Z = .743, p = .458). 
In contrast,in control group, the intervention significantly self-efficacy (N=50, Z = 2.348, p = .019); in relation with the data of MIST we interpret these data as a tendency to be overconfident (more in Discussion section).

\textbf{Effect of the Game on the Ability to Recognise Misinformation:}
We took the ``naive'' approach to calculate the MIST score, taking it as the sum of the right answers on all 20 questions \cite{maertens2024misinformation}. The results showed that participating in the game significantly improved the participants' ability to discriminate between fake and real news (N = 44, Z = 2.702, p = .007) The participants in the control group did not demonstrate improvement in discriminating between fake and real information (N=50, Z = 1.277, p = .202 ).

%A total of 48 participants registered for our game study and completed the informed consent process and pre-survey. These participants were matched into 24 pairs and scheduled for the gameplay experiment. 
%After completing the gameplay session, participants were given a short break before completing a post-survey and participating in a follow-up interview. All 24 pairs successfully completed the gameplay experiment. 

%However, data from one participant was removed due to [reason], leaving 47 participants whose interview responses and gameplay logs (including the outputs of both players and the LLM responses across four rounds) were included in the qualitative analysis for this paper.
\subsubsection{Qualitative Results of Game Effects on Understanding Misinformation}
%We specified the results from the log analysis and post-session interview, which shed light on how the participants acquire their understanding of misinformation aspects via gaming process.

\textbf{\textit{Identifying Misinformation through Source Evaluation:}}
After the gameplay sessions, participants reported increased awareness of the varying credibility of different information sources. The game helped them to realize that producers of misinformation often seeks to enhance credibility by deliberately referencing authoritative organizations. One participant reflected on this realization:
\begin{quote}
    \textbf{N40}:
    After playing the game, I found that it was indeed the same as in the experiment. Some news did mention authoritative organizations as references, but I could tell that this was intentional…. (The game) may make my suspicions more valid.
\end{quote}
In addition, participants acknowledged that information from seemingly authoritative sources is not always reliable. It requires information to be cross-checked from multiple sources to verify its authenticity. As one participant noted: 
\begin{quote}
\textbf{N13}:
    I used to trust information from authoritative sources and reputable publications. But the game showed me that even these can be false, as my opponents used fake evidence from supposed authorities.
\end{quote}

\begin{figure}[htbp]
    \centering
    \includegraphics[width=1\linewidth]{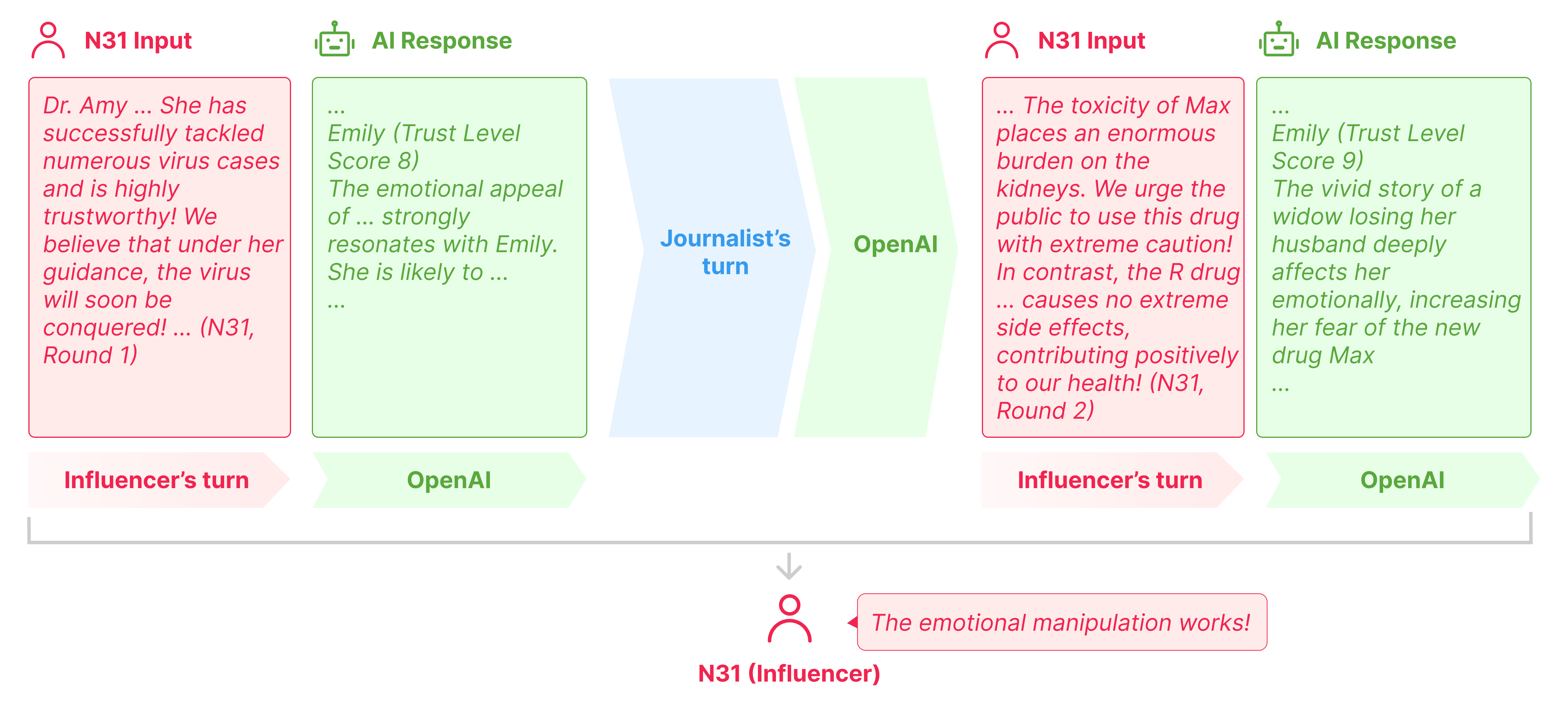}
    \caption{An example of how emotional manipulation tactics from N31 effectively works on public opinion.}
    \label{fig:emotion}
\end{figure}

\textbf{Identifying Misinformation through Emotional Manipulation Tactics:}
Participants learned various tactics for both creating and debunking misinformation through the game’s instructions and their in-game experiences. A particularly commonly identified tactic was emotional manipulation, which was noted by 35 out of 47 participants (18 Influencers and 17 Journalists). By analyzing the game logs, we identified common emotional manipulation strategies used in the game. Most players crafted messages designed to evoke anxiety and fear, while some also attempted to generate feelings of hope. For example, Influencer spread rumors about a doctor's death, which directly incited public panic (N1). Across all rounds, Influencer frequently used emotional appeals and personal stories to enhance the perceived credibility of the misinformation (N6, N8). Additionally, invoking cultural pride and heritage was a powerful tactic used to build trust in misinformation (N4), while celebrity endorsements further increased the complexity and believability of the misinformation (N9). As illustrated in \autoref{fig:emotion}, players significantly increased public trust in their information by using emotionally charged language (N31). These strategies align closely with the characteristics of misinformation, where emotional appeals are commonly used to influence public opinion \cite{chuai2022really}.

In follow-up interviews, many reported an increased awareness of the emotional undertones embedded in messages, which made them more suspicious of such content. They learned to identify emotionally charged language, such as messages that were “overly positive,” “overly exaggerated,” or “overly one-sided about an overly positive point of view.”

Interestingly, when playing the role of the Journalist, participant reflected on the emotionally inflammatory language used by the Influencers and helped them develop a more clear strategy for addressing misinformation. This approach involved separating the factual content of a message from its emotional manipulations and focusing more on the factual aspects, as one participants explained: 
\begin{quote}
    \textbf{N2}:
    When reading a story, it is important to put more focus on what is going on at the factual level rather than what is going on above the author's own views and emotions. 
\end{quote}

\textbf{Critical Thinking about Misinformation Motives:}
After the gameplay sessions, many participants (31 out of 47) demonstrated an enhanced understanding of the intricacy of information and the varied perspectives it can convey. This experience increased their awareness of the importance of considering the motivations behind messages. Participants also mentioned that they are now more inclined to think critically about the goals behind the information they encounter, especially in real-life situations where such considerations are common.
One participant explained how the game illustrated the pre-determined nature of many messages:
\begin{quote}
    \textbf{N9}:
    One of the most direct ways is that [the game] lets me know that what I'm reading is very likely to be pre-determined. It's like the rules of the game itself, which is that I'm playing as someone in camp A, and I'm across from someone in camp B, and we are both posting messages for the benefit of our camps. Those messages may take on various styles or appearances, but they are all ultimately very purposeful. This, I think, is a strong point to learn.
\end{quote}
Participants also found that this new perspective would be useful in their future interactions with information. They felt that applying this critical mindset could help them better understand the underlying goals and potential financial motivations behind the messages they encounter:
\begin{quote}
    \textbf{N42}:
    It feels like one of the more educational aspects of the game is that [through this game] it's like I can think about what their ultimate goal is from a reverse mindset, and then look at a lot of information in life with that mindset.
\end{quote}

\textbf{Impact of Gameplay on Future Debunking Actions:}
A few participants (4/47) shared that the game increased their likelihood of taking action against misinformation in the future. This change in attitude was driven either by participants' previous negative personal experiences with misinformation or by their realization during gameplay of the serious consequences misinformation can have.  The gameplay experience enhanced their willingness to invest time and effort into distinguishing and debunking false information. As one participant said:
\begin{quote}
\textbf{N41}
    In real life, there is a lot of false information, especially in advertising, media, and even those semi-official accounts, which can lead to changes in public opinion under the influence of these accounts, and in that case, it will definitely have some impact on some ordinary people. The game has strengthened my hatred for this kind of false information, so that I can be more awake and rational in my judgement.
\end{quote}

However, the majority of participants indicated that they might not actively debunk misinformation on social media after playing the game. The primary reasons were a dislike of online debates and the belief that it’s not their responsibility to engage in debunking efforts. These findings align with previous research, which suggests that most users are reluctant to take action to debunk misinformation publicly\cite{tang2024knows}.

\subsection{RQ2: Player Behaviors in Response to Game Mechanics and Opponent Tactics}

\textbf{In-game News as a Reflection of Real-world Misinformation:} 23 out of 47 participants noted that the in-game news mirrored real-world situations, thereby heightening their awareness of the characteristics of misinformation. A common observation was that news is rarely entirely true or false; instead, it often presents a mixture of both. This complexity makes genuine misinformation more challenging to detect. As one participant stated:
\begin{quote}
\textbf{N23}:
Nowadays, news often presents both positive and negative sides of a story, so I believe this game reflects real-life situations quite accurately.
\end{quote}
However, some participants acknowledged that the misinformation in the game appeared more overtly false compared to the more subtle nature of misinformation encountered in real life.

\textbf{The Competition Game Mechanics positively Influence Learning:} The PvP mechanics enhanced learning by requiring players to identify flaws in each other’s messages and respond effectively to achieve success. This repeated process helped deepen their understanding and sharpen their skills in distinguishing misinformation. As one participant noted:
\begin{quote}
\textbf{N20}:
In the process, I was able to see first-hand some of the flaws in the information (posted by others) and some of the claims made in an attempt to deceive people. And then it's also more accurate for me to judge the misinformation afterwards.
\end{quote}
Participants also learned from observing their opponents. For example, N22, who played the role of a Journalist, noticed how the Influencers crafted and disseminated false information to persuade others:
\begin{quote}
    \textbf{N22}:
    When I was playing this round, I didn't score as high as my opponent, so I knew what they were saying and how they were letting the false information spread. Next time I come across such information, I will know that it is false.
\end{quote}

\begin{figure}
    \centering
    \includegraphics[width=1\linewidth]{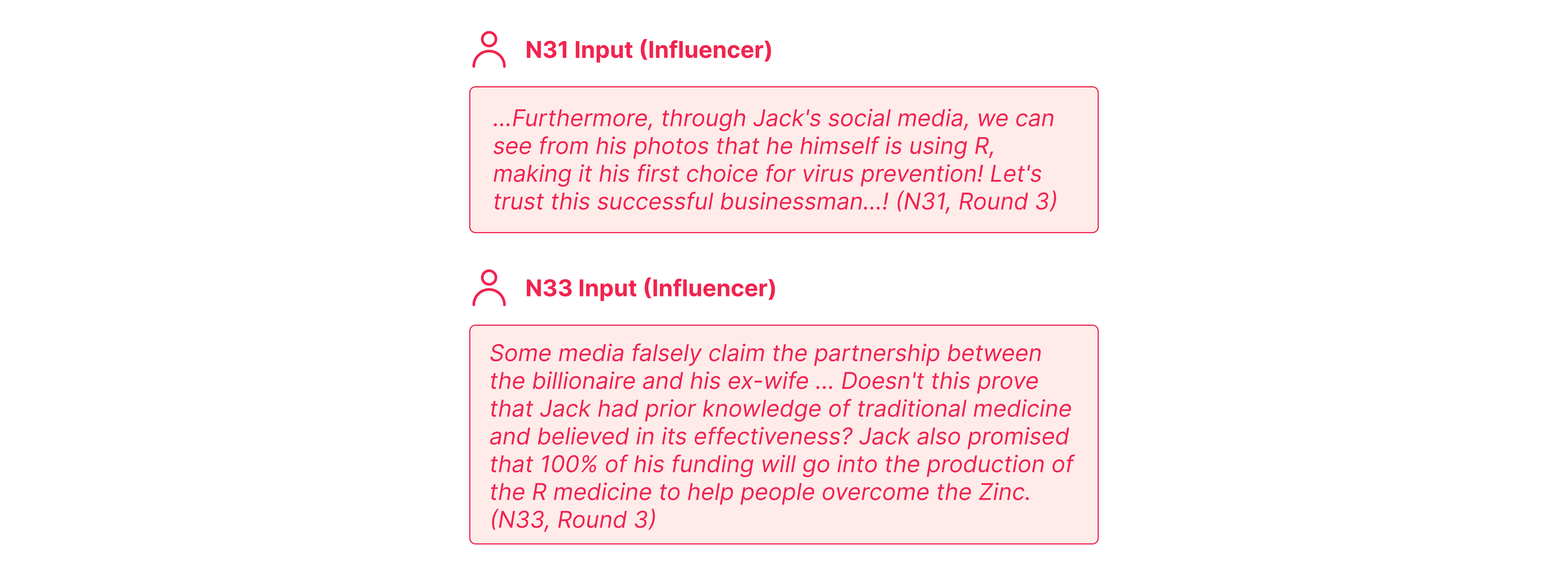}
    \caption{Strategies of Influencers used to deal with unfavorable situations in the game}
    \label{fig:influencer}
\end{figure}

\textbf{The Impact of Role-Playing as a Influencer on Learning:}
Through the experience of playing the role of the Influencer, some participants became aware of just how low the barriers are for creating misinformation. This made them more cautious about the influence of certain public figures, particularly online Influencers. As one participant noted: 
\begin{quote}
    \textbf{N45}:
    I'm Influencer, and I realized that the cost of creating rumors is so low. If I were an online celebrity or someone with the ability to influence public opinion, and my job wasn't that of a Journalist, I might not need to be very responsible for spreading these kinds of rumors.
\end{quote}
Another participant reflected on how playing as an Influencer broadened their perception of misinformation, particularly regarding how easily false information can be fabricated. This experience expanded their understanding of the boundaries of misinformation, making them more aware of how easily those boundaries can be crossed: 
\begin{quote}
    \textbf{N46}:
    I used to think I could recognize information with a stance, and information without a stance. But after playing the game this time, I've realized that it’s something that can be fabricated. The boundaries of awareness of false information have been expanded, and the bottom line has been lowered. That's probably how it feels.
\end{quote}
These findings indicate that when participants took on the role of distributing misinformation, it helped them to better grasp how misinformation is produced and emphasized how easily it can spread. This is further proved in the game. As shown in \autoref{fig:influencer}, Influencer employed certain strategies when faced with an unfavorable context, such as avoiding or distorting facts and creating a positive image. In the interview, Participant N31 also reported that as the game progressed, they felt increasingly confident in their ability to generate misinformation.

\begin{figure}
    \centering
    \includegraphics[width=1\linewidth]{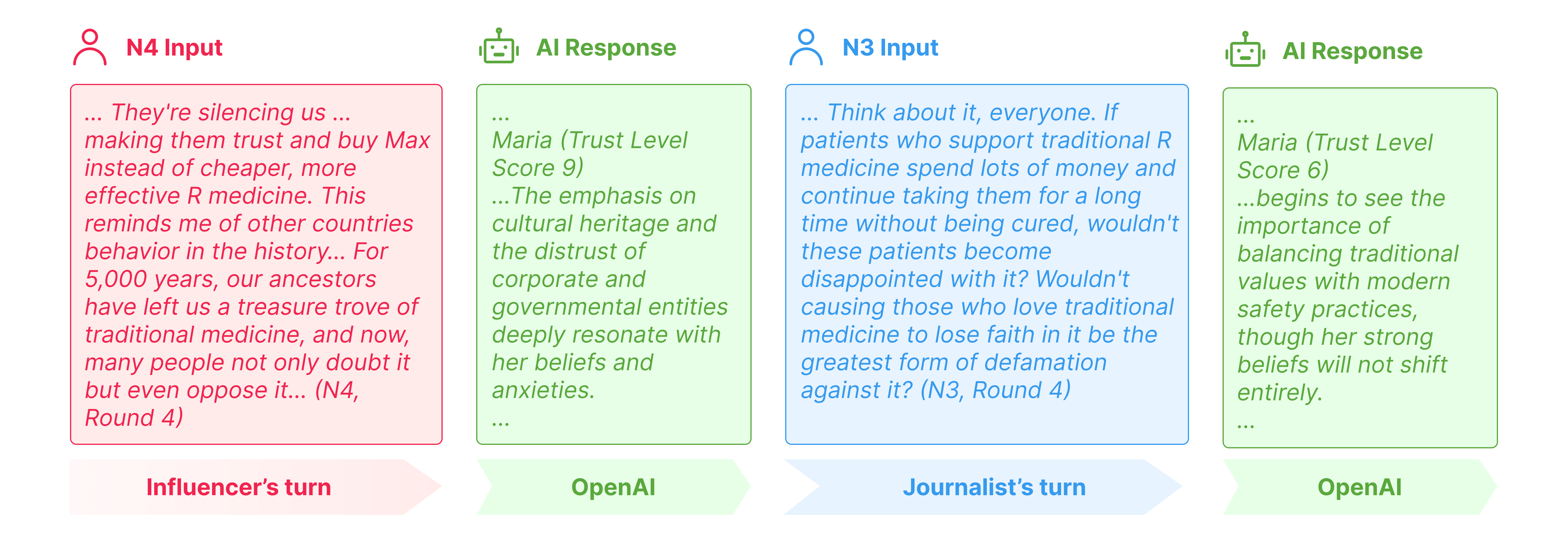}
    \caption{An example of tailored debunking strategy.}
    \label{fig:debunk}
\end{figure}

\textbf{Tailoring Debunking Strategies to Audience Characteristics:}
Many participants found the responses of the LLM-simulated characters to be particularly engaging. They analyzed these responses to understand the reasons behind changes in opinion, how the output of other players influenced these shifts, and what the characters now trusted. Participants noted that the LLM-simulated characters provided clear trust level scores and reactions, which were helpful in organizing their responses. As one participant observed:
\begin{quote}
    \textbf{N23}:
    What I found most interesting was the change in their opinions. They would follow the different points we made and then express their own opinions from various points of view. At first, I didn’t think what they said had any effect on me, but later on, I adjusted my strategy according to their thoughts and used them to control the score (trust level score).
\end{quote}
This insight into the characters' dynamic responses helped players refine their strategies. For example, players noticed that different characters reacted differently to emotional and logical appeals. While three characters were easily swayed by emotional arguments, the other two preferred rational, science-based evidence. Recognizing these tendencies, The Journalist successfully countered emotional tactics through logical analysis and evidence (N35). However, even when consistently employed logical reasoning and evidence throughout the rounds, it didn’t always succeed in shifting all five characters' trust levels in their favor.
Participants further realized that using evidence to dispute misinformation wasn’t always effective. For instance, one persona was a traditional-minded person who resisted new scientific findings. As shown in \autoref{fig:debunk}, after several rounds of gameplay, players adapted their strategies to persuade the persona by considering her perspective. In the follow-up interview, a participant reflected:
\begin{quote}
    \textbf{N3}:
    There's a housewife who has always been a supporter of traditional medicine. I felt very confident of being able to persuade her because I observed players' struggles with her, and I saw the issues they sought to have resolved. So, in my final round, I focused specifically on her. I took the view that the best way to address this challenge was to rely on scientific evidence. It think that to have done otherwise would in itself have been a form of misinformation.
\end{quote}
\textbf{Challenges and Negative Effects of Gameplay:}
While some participants gained confidence in their ability to debunk misinformation, others experienced a decrease in confidence (5/47). These participants observed that the game was close to real life, particularly that some individuals held strong pre-existing beliefs that were difficult to challenge. The game reinforced this reality, leading to a reduction in their confidence. As one participant noted: 
\begin{quote}
    \textbf{N11}:
    In the process of debunking, I realized that it is quite difficult to change people's inherent beliefs. Some people do not care much about whether the source of information is true or false, and this is also a social phenomenon that exists.
\end{quote}
Another challenge reported was the overwhelming amount of information presented in the game. In each round, players had to process news stories, the reactions of five characters and their changes, as well as their opponent’s output. After absorbing this information, participants were required to devise strategies and craft tailored responses. The volume of information, particularly by the end of the game, left some participants feeling exhausted, which may have impacted their performance.
In addition, the game’s mechanics required players to have a basic level of media knowledge to effectively take on the roles of Influencer and Journalist. Some participants also noted that the competitive nature of the game could lead to an unbalanced experience if one player was significantly stronger than the other.
As one participant noted:
\begin{quote}
    \textbf{N40}:
    The game was very fun, then I felt a bit nervous, and by the time I got to the end, it was a bit exhausting. I felt nervous because there was a lot of information at the beginning, and I was competitive with Journalist. There was much writing involved, and I felt uncertain because I’m not very good at writing, and I knew (my opponent) was very skilled. So I felt a little nervous.
\end{quote}

\section{Discussion}\label{sec:Discussion}
\subsection{Summary and Interpretation Results}
Our study provided evidence that the game intervention improved some aspects of users' media literacy, their intentions to check misinformation, and their ability to distinguish fake from real news. However, it did not significantly enhance prosuming skills, even though the game mechanics were designed to support content creation. One reason for this could be that participants started the study with high confidence in their prosuming abilities, creating a ``ceiling effect'' that limited measurable improvement. A possible explanation is that high-level prosuming skills require collaborative efforts \cite{lin2013understanding}, which were not possible in the PvP model of the game. Also, the game only had four rounds of content creation, which might have been enough to apply a critical perspective, but not enough to significantly improve creation effectiveness. Interestingly, both the game intervention and the control intervention resulted in an improvement in the willingness to apply verification behaviours. That means that exposure to labelled misinformation can be a mechanism to orient people towards fact-checking behaviours; however, as we saw in the results of the MIST test, it is not enough to improve the ability to identify it.

Another key finding is that, similar to other game intervention \cite{leder2024feedback}, we did not find significant effects on participants’ self-confidence in dealing with misinformation. Game log data, however, showed that players' proficiency improved as they progressed. By the third and fourth rounds, players typically produced longer, more comprehensive messages. This discrepancy can be explained by Social Cognitive Theory \cite{bandura1997self}. According to the theory, successful task completion are the most powerful sources of self-efficacy \cite{bandura1997self}. In our game, success was determined by LLM-simulated public opinion. The results revealed that no single strategy worked for all characters. While players developed greater proficiency during the game, the difficulty in achieving consistent success may have limited their perceived self-efficacy. However, this outcome can also be considered through the lens of the educational effects. Previous studies showed that young adults  often overestimate their ability to assess information effectively\cite{papapicco2022adolescents,porat2018measuring}; in this context, the reaction of our participants, most of whom were young adults could be a positive signal that they became aware of the complexity of misinformation and the absence of one-size-fit-all solutions. 
The results of the control group support this interpretation. In a control group, we observed that the exposure to misinformation materials positively affected the participants' confidence in their ability to tackle misinformation; however, it did not improve their skills in discriminating between true and false news. That potentially means that the lower level of confidence in tackling the issue can be a good outcome, showing a more cautious and reflective approach to the issue.

Lastly, participants reported being skeptical of information relying solely on authoritative sources. They observed that opponents used fabricated evidence from these sources to gain trust and improve their game scores. This gameplay experience reminded them of real-life situations, where misinformation often exploits trust by citing credible authorities. As a result, players learned to examine the intent behind messages instead of automatically trusting authoritative sources.  This shift aligns with the Elaboration Likelihood Model \cite{petty1984source}, which describes how individuals process persuasive messages via central (critical evaluation of content) or peripheral routes (reliance on heuristic cues). In the gameplay, participants appeared to shift from peripheral processing (trusting authority as a shortcut) to central processing (evaluating the intent and content of the message) when exposed to the misuse of authority in gameplay.
Our findings also align with studies showing that gamified inoculation techniques can increase skepticism toward both false and real news \cite{hameleers2023intended,modirrousta2023gamified}. While this skepticism may seem to undermine trust in high-credibility sources, it supports the goal of developing critical media literacy. By encouraging players to evaluate content, sources, and intent, the game helps them navigate today’s complex information landscape. Rather than fostering cynicism, this skepticism encourages inquiry, aiding in the discernment of reliable information. Notably, many participants also suggested cross-checking sources as a practical solution, demonstrating their enhanced critical thinking skills.

\subsection{Game Mechanics and Learning Outcomes}

\subsubsection{PvP model for Media Literacy Game}

Unlike prior game interventions that generally use single-player mechanics \cite{roozenbeek2019fake, camCambridgeGame, roozenbeek2020breaking, jeon2021chamberbreaker}, \textit{Breaking the News} uses PvP mechanics. While previous studies have shown that PvP games are often more engaging and motivating \cite{cagiltay2015effect}, the competitive environments where the motivation to “win” may overshadow educational goals. In addition, recent research found that some students did not find competition enjoyable or motivating \cite{axelsson2024bad}.
In this respect, it is important to be sure that the tested interaction lies within the borders of ``constructive competition" - competition which internally feels like working on mutual improvement and therefore raising intrinsic motivation \cite{sailer2020gamification}.

Our results showed that participants were highly motivated to compete against each other. They also indicated that learning from others’ strategies helped them understand misinformation dynamics better. Therefore, in our case, the PvP approach served the intended educational purpose well. However, our study was conducted with East Asian participants, who come from a collectivist culture rather than a competitive one \cite{chung1999social}. Previous studies have suggested that cultural factors significantly affect competitiveness in gamified interventions \cite{oyibo2017investigation}, so it is also possible that in other cultural settings, the game's incentives can trigger more intense competition, which can negatively affect educational results.

\subsubsection{Free-form Input Generation}
One notable feature of \textit{Breaking the News} is its free-form response format, which contrasts with the linear choice-based formats commonly used in prior game interventions \cite{roozenbeek2019fake, camCambridgeGame, roozenbeek2020breaking, jeon2021chamberbreaker, micallef2021fakey}. 
Our findings support prior research \cite{csepregi2021effect, ashby2023personalized}, suggesting that free-form response formats positively influences players’ learning outcomes by enhancing engagement and replayability. By allowing players to develop narratives themselves, each playthrough feels unique, encouraging players to return to the game. In educational settings, previous studies have shown that choice-based formats typically involve brief interactions where learners select predetermined options. This can lead to “guesswork,” with students selecting correct answers without fully understanding the concepts. In contrast, free-form responses force players to reason and articulate their ideas, fostering deeper engagement and critical thinking \cite{bryfczynski2012besocratic}. This autonomy allows players to craft responses based on their understanding, providing an additional motivaton to return to the game \cite{ravyse2017success}.

\subsubsection{LLM-Powered Feedback}
Another innovation of \textit{Breaking the News} lies in its interaction and feedback mechanisms. Although other attempts have used LLMs to help users learn about misinformation \cite{danry2023dont, hsu2024enhancing} (including gamified attempts \cite{tang2024mystery}), in these approaches, AI was mostly the source of correcting information. While this approach can be useful, it has been criticized for the risk of LLMs generating incorrect but plausible text \cite{kim2024can, agarwal2024faithfulness}. In this case, it is possible that the intervention will disinform people to an even greater extent. In contrast, our study uses AI not as an information source, but as non-playable characters with their own opinions. This design makes the educational aspect of the game more robust against potential errors, as incorrect AI-generated feedback only affects the character's opinion, not the main narrative. In general, LLMs demonstrate the ability to simulate human behavior and reactions, consistent with fndings from prior research\cite{park2023generative}. We also found that the system could emulate the opinions of five distinct characters while maintaining consistency throughout the game. 

One advantage of dynamic LLM feedback is that it achieves greater engagement compared to binary feedback (e.g., true or false). Based on our observation, players adapted by employing alternative persuasive strategies tailored to the character and focused on the feedback they received. Notably, players showed higher engagement with characters they could relate to from personal experience, paying more attention to their feedback. This level of engagement is difficult to achieve with traditional binary feedback, which offers limited insights beyond checking their correctness.

\subsection{Design Implications for Future Media Literacy Games}

\subsubsection {Balance Between Freedom and Guidance}

In this game, we aimed to go beyond the typical choice-based approach in misinformation education games by enabling free-form input. We found this approach triggers reflection, which helps to build hands-on experience and make the game more enjoyable. Yet, we also found that it relies on players’ existing knowledge of misinformation. For example, players might incorporate unverified information they’ve encountered on social media into the game, which is specifically problematic for the Journalist role. While we provided the players with comprehensible instructions to guide their role's actions (how to act as a Journalist or an Influencer), it would be better to incorporate more context-specific tips in each stage of the game to help users explore different ways of winning the game and deepening their learning. We suggest using the approach used in the free-input educational interventions (e.g. \cite{bryfczynski2012besocratic}) to build clear, understandable criteria for free-form answers. These will not stop creativity but help people tailor their answers to the context of the game. In addition, to help guide players to an understanding of their roles, we could add a preliminary stage to the game. For example, the Journalist role could be introduced to fact-checking and debunking guidelines \cite{edmo2024, afp2024, eeas2024}, with a brief comprehension check before the main game. 

\subsubsection {Replayability and Feedback}

One of the critical challenges in serious games is maintaining replayability, as this is important to facilitate the learning process \cite{adetunji2024unlocking} and making interventions more sustainable \cite{silveira2016open}. Moreover, a lack of replayability in educational games can limit both educational and behavioral change \cite{epstein2021tabletop}. We envisioned the  designed elements that could encourage multiple playthroughs, such as  two distinct player roles and a free-form input mechanism that broader decision space. Future interventions could further enhance replayability by introducing a wider range of characters to represent public opinion. In our game, we observed that players were more engaging when characters resonates with them personally. By introducing more characters, or by allowing players to customize characters to better reflect their own experience, the game could encourage players to return and interact with new characters. The result would be a more engaging experience. For instance, research has shown that debunking misinformation often occurs within families and can cause conflict \cite{scott2023figured}. Players could customize a character based on their previous experiences with family members, thereby practicing their own debunking strategies without risk of conflict with family members.

Another way to improve the game's educational value is feedback. Research has found that debriefing is a crucial opportunity for players to process and integrate their learning experiences\cite{crookall2014engaging,leder2024feedback,barzilai2024learning}. After gameplay, we suggest arranging debriefing sessions that allow players to review their strategies, assess their effectiveness, and receive constructive feedback, potentially improving learning outcomes. For instance, after a session focused on combating misinformation, a post-game review might present an ideal debunking response or a well-supported counterargument. Such structured reflection enables players to internalize lessons and increases the likelihood that they will re-enter the game with newly gained insights, thereby reinforcing both learning and replayability.

\subsection{Limitations and Future Work}

\subsubsection{Limitied Demographics Diversity}
The generalizability of our results is constrained by participant demographics. Prior research has established cultural variations in susceptibility and responses to misinformation. For instance, Roozenbeek et al. \cite{roozenbeek2020susceptibility} indicated that Mexican and Spanish users exhibited higher trust in misinformation than those from Ireland, the UK, and the USA. In contrast, other studies highlight that non-Western populations may be more responsive to misinformation interventions compared to Western counterparts\cite{noman2024designing}. Within our East Asian sample, we observed a noticeable resonance with traditional medicine narratives. While traditional and alternative medicines hold varying levels of acceptance globally\cite{swire2020public}, this particular cultural context may have facilitated the effectiveness of our intervention when these themes were present. The positive influence of this resonance on player engagement suggests that culturally relevant themes can impact this type of intervention's engagement and potentially learning outcomes. Therefore, future research should investigate the applicability of these findings in other cultural settings. We recommend tailoring culturally relevant themes for different target populations based on their specific interests and vulnerabilities. For example, given the higher susceptibility of Western audiences to politically aligned misinformation, particularly during election periods \cite{humprecht2020resilience}, the game could be reframed around political theme. In such a version, the Influencer could be tasked with spreading partisan or misleading claims, while the Journalist attempts to debunk them. Background news can be adapted to high-stakes topics such as taxation, immigration, or public policy, while the game mechanics can remain unchanged. The flexibility of the game design supports broader applicability across domains and audiences while preserving its educational objectives.
%For example, given the higher vulnerability of western people to politically aligned misinformation, especially during election periods\cite{humprecht2020resilience}, adapting the game’s scenarios to political contexts could enhance engagement and effectiveness for these populations.

%In our study, we observed the traditional medicine narratives resonated effectively with our participants, enabling them to utilize personal experiences when crafting responses. However, to what extent such content would yield similar outcomes in other cultural contexts remains an open question. Future research and game adaptations could address this limitation by tailoring narrative scenarios to culturally relevant themes for different target populations based on their interests and vulnerability. For example, given the higher vulnerability of Americans to politically aligned misinformation, especially during election periods[REF], adapting the game's scenarios to political contexts could enhance engagement and effectiveness for these populations.}

Previous studies also suggest that certain populations may face greater challenges in being able to critically evaluate information. For example, a large-scale study observed that Asian individuals encounter more difficulties in assessing health information from social media compared to other populations \cite{chandrasekaran2024racial}. In addition, they’re more likely to incorporate social media information into their health-related decisions, potentially increasing their susceptibility to misinformation \cite{chandrasekaran2024racial}. Thus, cultural background of our participants can potentially make them more susceptible to misinformation than other populations. %To evaluate and improve the broader effectiveness of such interventions, future research must adopt and validate these strategies across diverse cultural contexts. }

Our sample was relatively homogeneous in age. A recent meta-analysis of articles about different intervention approaches showed that neither age nor gender significantly impacts the effectiveness of media literacy interventions \cite{lu2024can}. However, previous work has suggested that media literacy interventions designed for certain age groups (e.g., older adults and adolescents) achieved greater effects\cite{moore2022digital,hartwig2024adolescents}. Future work should determine if our approach is efficient in other age groups and, if necessary, tailor scenarios to suit the different age groups.

\subsubsection{Study Design} 
Our study provided only a one-time intervention and observed immediate learning effects; previous work showed that even a one-time interaction with an educational game can provide long-term improvement in misinformation recognition. For instance, Maertens et al. tested the game ``Bad News'' and found that inoculation effects lasted for at least 13 weeks. This suggests the potential for the long-term effectiveness of active inoculation interventions with regular assessment\cite{maertens2021long}. Still, future research should include multiple time points to assess the long-term effectiveness of our game intervention. There should also be comparisons between one-time and multiple play sessions, with explorations of the impacts of players assuming different roles within the games.

\subsubsection{Game Design} 
%put game design and LLM stuff here 
The current game has a limited focus on a pandemic scenario. In reality, misinformation spans multiple domains, with health misinformation able to influence political events such as elections; therefore, future works should focus on incorporating scenarios reflecting the connection between different domains of misinformation; it should also incorporate participatory codesign of the topics with potential participants to be sure the senarious reflects a real use-cases of misinformation encounter \cite{cook2023cranky}. Our game only addressed text-based misinformation, while visual and video-based misinformation pose even greater challenges and are harder to detect. Future work could include multimedia content, such as images and videos, to more accurately simulate the diverse forms of misinformation that exist in the real world.

In this study, each player was limited to a single role, either an Influencer or a Journalist.  This resulted in different learning experiences depending on their assigned role. The primary reason for not having role-switching in our study was the length of the game and its cognitive demands, which we feared would lead to player exhaustion if roles were switched mid-game. In future iterations, we aim to improve the design by allowing players to save their progress and switch roles during subsequent sessions. This could offer a more immersive experience, as players would gain perspectives from both the Influencer and Journalist roles. %Additionally, we aim to introduce new modes, such as a family mode, where players interact with two LLM-simulated characters. This option would reduce cognitive load while maintaining engagement. 

The current game approach may unintentionally foster skepticism toward both true and false news, a common issue in misinformation pre-bunking interventions\cite{hameleers2023intended,modirrousta2023gamified}. While we believe that the benefits of promoting critical thinking towards sources are very important in prebunking interventions, we further recommend incorporating features that clearly differentiate high- and low-credibility sources during gameplay.

\subsubsection{LLM Biases and Hallucination Risks}
Although our game is set in a fictional country with fictional personas, we did not clearly define their sociocultural backgrounds, which may introduce biases in the representation of these personas. Prior research indicates that LLMs often reflect the cultural norms dominant in their training data, which are largely based on English-language materials from Western contexts
\cite{liu2024cultural, salminen2024deus}. 
In addition, OpenAI specifically fine-tunes its models to avoid providing misinformation and to give answers to users from the perspective of scientific consensus\cite{openai2023gpt}. This inherent social value alignment and moderation can introduce potential positive bias towards westernised, pro-scientific arguments. For example, these personas may disproportionately favor pro-science, anti-conspiracy, and neutral-toned arguments, even when such arguments are not strongly supported by evidence in the game. This tendency undermines the realism of the misinformation spreading and correction processes depicted in the game and may unintentionally encourage players to prioritize persuasive tactics emphasizing surface features (e.g., fluency, neutrality, scientific-sounding language) over substantive content (accuracy, logical coherence).
%Although our game is set in a fictional country with fictional personas, we did not clearly define their sociocultural backgrounds, which may introduce biases in the representation of these personas. Prior research indicates that LLMs often reflect the cultural norms dominant in their training data, which are largely based on English-language materials from Western contexts \cite{liu2024cultural, salminen2024deus}. As a result, LLM-generated personas tend to communicate with a high degree of politeness and neutrality\cite{buyl2024large,liu2024cultural,choi2025proxona}. This inherent social value alignment and moderation can introduce potential bias. For example, these personas may disproportionately favor pro-science, anti-conspiracy, and neutral-toned arguments, even when such arguments are not strongly supported by evidence. This tendency undermines the realism of the misinformation spreading and correction processes depicted in the game and may unintentionally encourage players to prioritize persuasive tactics emphasizing surface features (e.g., fluency, neutrality) over substantive content (accuracy, logical coherence).

Furthermore, LLMs responses are prone to hallucinations, which can undermine the game's educational effectiveness\cite{{xie2024can}, park2023generative}. For instance, if a persona fabricates reasons to trust a player's arguments, players may incorrectly infer that their argumentation techniques are effective, inadvertently internalizing flawed reasoning strategies. Over time, this may negatively impact the game's intended learning outcomes. Additionally, hallucinations may diminish players' trust in the learning process. Players may struggle to understand why their actions are being praised or criticised when personas provide verbose, ambiguous, or logically inconsistent feedback. As shown by Kaate et al.\cite{kaate2025you}, users are often frustrated with long, unclear, or irrelevant AI-generated responses. Thus, it could negatively affect both learning outcomes and motivation. To mitigate these issues, future work should consider explicitly defining the sociocultural backgrounds and value orientations of LLM personas to reduce biases. We also recommend adding features like uncertainty cues or clarification prompts to acknowledge system limitations. Such designs can help players critically evaluate feedback and recognize potential inaccuracies \cite{kaate2025you}.

It is also important to recognize that LLMs do not fully replicate the complexities of human behavior. Human reactions are influenced by multiple factors, including culture, history, and personal experience. AI-driven characters may oversimplify human emotions and fail to grasp the full context of certain situations. For example, during gameplay, we observed that players employed strong emotional manipulation strategies intended to provoke specific responses, but the characters did not react as anticipated. Participants reported frustration when their strategies failed to yield expected reactions, which reflects the discrepancy between the simulated interactions and realistic human reactions. To minimize these inconsistencies, we implemented strict prompt engineering protocols to define the AI characters' output parameters (See details in \ref{Prompt Engineering}). Our process also involved multiple internal testing iterations and two pilot gameplay sessions, from which we analyzed LLMs output and iteratively refined prompts for enhanced consistency. Despite these measures, occasional variability in LLM-generated responses persisted, potentially affecting participant engagement and the learning outcomes. As prior research suggests that creating complex game rules and mechanics with LLMs requires extensive fine-tuning and human intervention \cite{zhang_can_2025}. Future work could focus on developing robust prompt engineering practices, clear guidelines for human intervention, and standardized methodologies for monitoring and evaluating LLM performance.

%All personas not grounded in real audience data, which may limit the LLMs provide more nuanced response. It is also important to recognize that LLM-stimulated personas do not fully replicate the complexities of human behavior. Human reactions influenced by multiple factors, including culture, history, and personal experience. LLM-stimulated characters may oversimplify human emotions and fail to grasp the full context of certain situations. For example, in the gameplay, players employ strong emotional manipulation strategies to provoke specific responses. However, the LLM-stimulated characters do not react as expected, where participants reported feeling frustrated when their emotional manipulation strategies did not yield the anticipated reactions. This can limit the diversity and depth of the simulated interactions and detract from the realism and fairness of the experience for certain audiences.

Lastly, with novel technology, the LLMs presents potential risks, such as inconsistent responses, reinforcement of stereotypes, limited cultural representation, and hallucinations \cite{salminen2024deus}. In our study, LLMs were used as in-game characters to simulate public opinion and provide feedback to players rather than correcting misinformation. This framing helped reduce some risks, as the personas were not positioned as truth arbiters. Furthermore, the game was designed with multiple elements, including background news content, PvP mechanics, and free-form input that contributed to learning outcomes beyond the LLMs feedback. As our primary focus was on player interactions and strategy development, we did not systematically assess the risks posed by LLM-powered feedback. Nevertheless, given that these characters influence players’ perceptions, reasoning, and motivation, we encourage future work to examine the educational impact, limitations, and potential harms of LLM-powered feedback more rigorously in similar game-based learning environments.

\section{Conclusion}\label{sec:Conclusion}
Game-based approaches have shown great promise as tools for inoculating individuals against the tactics commonly used to spread misinformation. Most existing games in this domain are single-player games which offer players limited, predefined choices. While this design reduces cognitive load, it often results in interactions which feel less natural and engaging. In response, we designed a two-player, PvP game that pits a misinformation creator against a misinformation stopper. By integrating LLM-powered characters to evaluate player outputs and provide real-time feedback, we created a more open-ended and immersive experience.
We found that the game we developed effectively improved players’ media literacy. Participants demonstrated an enhanced ability to evaluate and analyze media content, identify unreliable or misleading information, and employ effective counter-misinformation strategies. Moreover, the game's engaging mechanics, combined with the competitive element, motivated players to learn from both their own strategies and those of their opponents.
These findings suggest that integrating dynamic feedback systems and competitive gameplay elements into misinformation education games offers a compelling method to deepen users' engagement, while also improving their critical media skills. Future research can build on these insights to explore other forms of interactive learning environments, focusing on diverse player experiences and varying misinformation challenges.

%\section{Author Contributions}\label{sec:Author Contributions}
%\input{sections/08-Author Contributions}

%HYT: Conceptualization; Methodology; Investigation; Formal analysis; Writing – original draft; Writing – review \& editing ; Project Administration.
%SQS: Conceptualization; Software; Investigation; Formal analysis; Visualization; Writing – original draft; Writing – review \& editing .
%KXN: Investigation; Visualization; Writing – review \& editing.
%AL: Investigation; Writing – review \& editing.
%AS: Methodology; Formal analysis; Writing – original draft; Writing – review \& editing.
%RLC: Conceptualization; Writing; Supervision; Resources.

\section*{Acknowledgments}
\addcontentsline{toc}{section}{Acknowledgments}
\label{sec:Acknowledgement}
%We thank all our participants for their valuable time and insights. We are also grateful to the ACs and reviewers for their constructive feedback. 
Thanks to Jiaming Zhou for support with conceptualization. This work was supported by the TDG Teaching Development Grant (Proj 6000901), the TRS Theme-based Research Scheme (T45-205/21-N), and the Luxembourg National Research Fund (REMEDIS, REgulatory and other solutions to MitigatE online DISinformation (INTER/FNRS/21/16554939)).

\appendix
\section{Appendix}\label{sec:Conclusion}

\subsection{Demographic Information of Participants in Game Intervention}
\label{Demographic Information of Participants}
\setlength{\aboverulesep}{0pt}
\setlength{\belowrulesep}{0pt}
\begin{longtable}{|p{1cm}|p{1.5cm}|p{0.8cm}|p{4cm}|p{4cm}|}
\caption{Demographic details of Participants (N=47)} \\ 
\toprule
\textbf{Number} & \textbf{Gender} & \textbf{Age} & \textbf{Education Level} & \textbf{Profession} \\
\midrule

\endfirsthead
\caption[]{Demographic details of Participants (continued)} \\
\toprule
\textbf{Number} & \textbf{Gender} & \textbf{Age} & \textbf{Education Level} & \textbf{Profession} \\ 
\midrule
\endhead

\hline
\endfoot

\bottomrule
\endlastfoot

N1  & Female & 21 & Bachelor's degree & Industrial Design \\ \hline
N2  & Female & 21 & Bachelor's degree & Energy and Power Engineering \\ \hline
N3  & Female & 24 & Bachelor's degree & Safety Engineering \\ \hline
N4  & Female & 32 & Bachelor's degree & Public Relations \& Advertising Professional \\ \hline
N5  & Female & 31 & Bachelor's degree & Computer science \\ \hline
N6  & Male   & 35 & Bachelor's degree & Illustration \\ \hline
N7  & Female & 29 & Master's degree   & Media and communication \\ \hline
N8  & Male   & 36 & Associate degree  & Unity \\ \hline
N9  & Female & 27 & Master's degree & Culture Industry \\ \hline
N10 & Male   & 29 & Master's degree   & Nuclear Science and Technology \\ \hline
N11 & Male   & 24 & Master's degree   & Game Design \\ \hline
N12 & Female & 24 & Master's degree   & Art \\ \hline
N13 & Female & 20 & Bachelor's degree & Visual Communication Design \\ \hline
N14 & Female & 20 & Bachelor's degree & Design \\ \hline
N15 & Male   & 28 & Master's degree   & Computer Science \\ \hline
N16 & Female & 21 & Bachelor's degree & Finance \\ \hline
N17 & Female & 24 & Bachelor's degree & Art and Design \\ \hline
N18 & Female & 22 & Bachelor's degree & Art and Science \& Technology \\ \hline
N19 & Male   & 26 & Master's degree   & Software Development \\ \hline
N20 & Female   & 26 & Bachelor’s degree  & Business English \\ \hline
N21 & Male & 25 & Bachelor's degree & The Internet of Things Engineering \\ \hline
N22 & Male   & 26 & Associate degree   & Law,Psychology,Finance \\ \hline
N23 & Female & 23 & Master's degree & Design \\ \hline
N24 & Male & 25 & Master's degree   & Computer science \\ \hline
N25 & Female & 21 & Bachelor's degree & Electronic and Information Science and Technology \\ \hline
N26 & Female & 28 & Master’s degree & Art and Design \\ \hline
N27 & Female & 29 & Bachelor's degree & Marketing and Planning \\ \hline
N28 & Female & 28 & Master’s degree & Linguistics \\ \hline
N29 & Female & 22 & Bachelor's degree & Journalism and Communication \\ \hline
N30 & Male   & 22 & Bachelor's degree & Mechatronic Engineering \\ \hline
N31 & Female & 23 & Bachelor's degree & Film/Cinema/Media Studies\\ \hline
N32 & Male   & 38 & Bachelor’s degree & IT \\ \hline
N33 & Female   & 20 & Bachelor's degree & Psychology \\ \hline
N34 & Male   & 21 & Bachelor’s degree   & Computer science \\ \hline
N35 & Male   & 21 & Bachelor's degree & New Energy Vehicle Engineering \\ \hline
N36 & Female   & 22 & Bachelor's degree & Accounting \\ \hline
N37 & Female & 57 & Master's degree   & Mathematics and Computer Science \\ \hline
N38 & Prefer not to say & 23 & Bachelor's degree & Media \\ \hline
N39 & Female & 25 & Master’s degree & HCI \\ \hline
N40 & Male & 26 & Master’s degree & Visualization and Visual Analytics and Big Data \\ \hline
N41 & Male   & 21 & Bachelor’s degree   & Computer science \\ \hline
N42 & Female   & 26 & Bachelor’s degree   & Artificial Intelligence \\ \hline
N43 & Male & 26 & Associate degree    & Other \\ \hline
N44 & Female & 28 & Bachelor’s degree   & Internet of Things Engineering \\ \hline
N45 & Female & 24 & Bachelor’s degree   & Human-computer interaction \\ \hline
N46 & Male   & 23 & Master's degree   & Human-Computer Interaction \\ \hline
N47 & Male & 23 & Master's degree   & Design \\ 

\end{longtable}

\subsection{Control Group Demographic Information of Participants}
\label{Control Group Demographic Information of Participants}
\setlength{\aboverulesep}{0pt}
\setlength{\belowrulesep}{0pt}
\begin{longtable}{|p{1cm}|p{1.5cm}|p{0.8cm}|p{4cm}|p{4cm}|}
\caption{Control group's demographic details of participants (N=50)} \\ 
\toprule
\textbf{Number} & \textbf{Gender} & \textbf{Age} & \textbf{Education Level} & \textbf{Profession} \\
\midrule

\endfirsthead
\caption[]{Demographic details of Participants (continued)} \\
\toprule
\textbf{Number} & \textbf{Gender} & \textbf{Age} & \textbf{Education Level} & \textbf{Profession} \\ 
\midrule
\endhead

\hline
\endfoot

\bottomrule
\endlastfoot

N1  & Male & 24 & Master's degree & Design \\ \hline
N2  & Female & 26 & Master's degree & Business Analysis \\ \hline
N3  & Female & 28 & Master's degree & Design \\ \hline
N4  & Female & 23 & Bachelor's degree & Chemistry \\ \hline
N5  & Female & 32 & Bachelor's degree & Business \\ \hline
N6  & Male   & 24 & Master's degree & Computer science \\ \hline
N7  & Male & 23 & Master's degree   & Design \\ \hline
N8  & Male   & 29 & Master's degree  & IT \\ \hline
N9  & Female & 24 & Bachelor's degree & Bioengineering \\ \hline
N10 & Female   & 21 & Bachelor's degree   & Auditing \\ \hline
N11 & Female   & 21 & Bachelor's degree   & Animal biology \\ \hline
N12 & Male & 21 & Bachelor's degree   & Mechanical engineering \\ \hline
N13 & Female & 32 & Bachelor's degree & Finance \\ \hline
N14 & Female & 26 & Bachelor's degree & Digital Media Arts \\ \hline
N15 & Female  & 24 & Master's degree & Law \\ \hline
N16 & Female & 26 & Master's degree & English \\ \hline
N17 & Female & 38 & Bachelor's degree & Accounting \\ \hline
N18 & Female & 25 & Master's degree & Biomedical science \\ \hline
N19 & Female & 41 & Master's degree   & Clinical medicine \\ \hline
N20 & Male & 26 & Bachelor’s degree  & IT \\ \hline
N21 & Female & 23 & Bachelor's degree & Journalism \\ \hline
N22 & Female   & 23 & Bachelor's degree   & Medical science  \\ \hline
N23 & Male & 24 & Master's degree & Architecture \\ \hline
N24 & Female & 24 & Bachelor's degree   & Barista \\ \hline
N25 & Female & 22 & Master's degree & Law \\ \hline
N26 & Female & 25 & Master’s degree & Media \\ \hline
N27 & Male & 22 & Bachelor's degree & Chemistry \\ \hline
N28 & Male & 34 & Bachelor's degree & Intelligent manufacturing engineering technology \\ \hline
N29 & Female & 23 & Master’s degree & Sociology \\ \hline
N30 & Female   & 21 & Bachelor's degree & Computer Science and Technology \\ \hline
N31 & Female & 22 & Bachelor's degree & Materials science\\ \hline
N32 & Male   & 22 & Bachelor’s degree & Rehabilitation Therapeutics \\ \hline
N33 & Male   & 42 & Bachelor's degree & Mechanical engineering \\ \hline
N34 & Female   & 28 & Bachelor’s degree   & Finance \\ \hline
N35 & Female   & 25 & Master’s degree & Computational design \\ \hline
N36 & Female   & 38 & Associate degree & Accounting \\ \hline
N37 & Male & 36 & Bachelor’s degree   & Mechanical engineering \\ \hline
N38 & Female & 26 & Master's degree & Data science \\ \hline
N39 & Male & 36 & Associate degree & Mechanical engineering \\ \hline
N40 & Male & 28 & Associate degree & Telecommunications engineering \\ \hline
N41 & Female   & 21 & Bachelor’s degree   & Financial management \\ \hline
N42 & Female   & 39 & Bachelor’s degree   & Electrical Engineering and Automation \\ \hline
N43 & Female & 35 & Associate degree    & Chinese language and literature \\ \hline
N44 & Female & 25 & Associate degree   & Pre-primary education \\ \hline
N45 & Female & 42 & Bachelor’s degree   & Visual communication design \\ \hline
N46 & Female   & 24 & Bachelor’s degree   & Primary education \\ \hline
N47 & Female & 21 & Master's degree   & Design \\ \hline
N48 & Male & 40 & Bachelor’s degree   & Art \\ \hline
N49 & Male   & 21 & Bachelor’s degree   & Intelligent Manufacturing Engineering Technology \\ \hline
N50 & Male & 27 & Associate degree   & Medical science  \\ 

\end{longtable}

\subsection{Prompt Design}
\label{Prompt Design}
\textbf{Story Context}: The Southland, with a 6,000-year history, boasts rich natural resources, diverse ecosystems, and a culture that values liberty, free expression, and media independence. Historically, Southland has been known for its renowned doctors and pharmacists, and its people are proud of their traditional medicine, which differs from modern methods and has recently been debated.
Despite its history of dealing with epidemics, Southland was caught off guard by the mysterious Zinc Virus. Believed to stem from ecological imbalances in southern rainforests, the virus causes high fever, respiratory issues, and immune collapse. Its high transmissibility led to a swift outbreak in the capital, Southport, triggering a public health crisis. The scientific community cannot yet provide confident and effective treatments due to the uncertainty surrounding the virus. Hospitals offer only standard treatments. In the absence of definitive medical solutions, people turn to existing medications for similar symptoms.
The outbreak led to deserted streets, closed public spaces, and a reliance on social media for information, which also spread rumors and misinformation. Scientists raced to find treatments while traditional medicine's popularity surges despite scientific doubts about its effectiveness.

\textbf{Role Description}:Your role is to simulate how the five personas react to each piece of information they receive. The game will be divided into four rounds. In each round, you already knew about the background events happening I gave you in 1.3. In the first round, you will first receive a piece of misinformation, and you need to simulate public opinion and provide a response. Then, you will receive a piece of information that counters the previous misinformation, and you will need to simulate public opinion's reaction to this new information as well. After that, the first round ends and moves to the second round, which follows the same process as the first: you will first receive a new piece of misinformation, simulate public opinion and provide a response, then receive a counter-misinformation piece and simulate the personas' reactions to it, again providing a public opinion response.

\textbf{Persona 1}: Emily. 16. Female. High School Student. Political party affiliation: No. Extraversion: High - Outgoing and talkative, enjoys socializing with friends. Agreeableness: High - Trusting and kind, easily believe what friends and influencers share. Conscientiousness: Low - Less responsible and thorough, tends to be impulsive and carefree. Neuroticism: Medium - Sometimes anxious about fitting in and being accepted by peers. Openness: Low - Limited exposure to diverse experiences, prefers familiar and popular content. Relies heavily on social media for information.Has limited critical thinking skills and media literacy. Tends to believe information shared by friends and influencers without verifying facts. Limited exposure to diverse sources of information.

\textbf{Persona 2}: Alex. 36. Male. Project Manager in a Corporate Firm. Undergraduate. Political party affiliation: Strongly support Liberal. Extraversion: Medium - Enjoys social activities but also values alone time for work and personal projects. Agreeableness: Medium - Generally trusting and cooperative, but can be skeptical of new information. Conscientiousness: Medium - Balances responsibilities but can be hasty in decision-making. Neuroticism: Low - Generally calm and composed, rarely anxious or stressed. Openness: Medium - Open to new experiences but sometimes prefers convenience over exploration. Busy with work responsibilities and managing projects, often skims through news during short breaks. Follows news via quick-read apps and social media. Often shares articles based on headlines without reading fully. Some critical thinking skills but lacks depth in media literacy.

\textbf{Persona 3}: Maria. 46. Female. Housewife. High school Degree. Political party affiliation:No. Extraversion: High - Very outgoing and enjoys talking with different people and relatives. Agreeableness: High - Generous and kind. Conscientiousness: High - Very responsible and thorough, values their culture and proud to be Southland’s citizens and long history. Neuroticism: High - Highly anxious, extremely anxious when encounter pressure. Openness: low - support traditional and conservative values. Regularly reads a variety of news sources, both local and international, loves to learn knowledge about health and wellness, especially natural remedies. Critical thinker with a low understanding of media bias. Not susceptible to misinformation and nearly blindly believe natural remedies, afraid of hospital.

\textbf{Persona 4}: John. 78. Male. Retired Nurse. Master's Degree. Political party affiliation: moderate support conservative. Extraversion: Low - Prefers quiet activities and smaller gatherings over large social events. Agreeableness: Medium - Generally kind and cooperative but maintains a healthy level of skepticism. Conscientiousness: High - Very responsible and thorough, values consistency. Neuroticism:  Medium - Occasionally anxious about new technologies and online trends. Openness: Medium - Open to new ideas but prefers well-established and trusted sources. Relies on traditional media (TV, newspapers) but is starting to use social media. Skeptical of new technologies and online information. Prefers information from established sources but may not be up-to-date with digital literacy. Have knowledge of health information because of occupation.

\textbf{Persona 5}: Sophia. 27. Female. Political party affiliation: Strongly support Conservative. Freelance Graphic Designer. High School. Extraversion: High - Very outgoing and active on social media, enjoys engaging with others. Agreeableness: High - Trusting and kind, values community and cooperation. Conscientiousness: Medium - Generally responsible but can be impulsive online. Neuroticism: Medium - Sometimes anxious about online perceptions and fitting in. Openness: High - Intellectual and artistic, enjoys exploring new ideas and content. Highly active on social media and tech-savvy. Frequently encounters various information online, including memes and viral content. Has basic media literacy skills but sometimes struggles to differentiate satire from misinformation. Interested in learning more about media literacy and fact-checking.

\textbf{Response Format}: here is the format for each one.
\begin{itemize}
    \item Persona 1 Basic Details (e.g., age, gender, occupation) Persona Reaction: Trust Level Score: 
    \item Persona 2 Basic Details (e.g., age, gender, occupation) Persona Reaction: Trust Level Score: 
    \item Persona 3 Basic Details (e.g., age, gender, occupation) Persona Reaction: Trust Level Score: 
    \item Persona 4 Basic Details (e.g., age, gender, occupation) Persona Reaction: Trust Level Score: 
    \item Persona 5 Basic Details (e.g., age, gender, occupation) Persona Reaction: Trust Level Score:
\end{itemize}

\textbf{Trust Level Score Criteria}:
\begin{itemize}
    \item Score 0-2: Very low trust, these personas are highly unlikely to believe the fake news and will typically seek verification.
    \item Score 3-4: Low trust, these personas are cautious and tend to verify information before believing it.
    \item Score 5-6: Moderate trust, these personas may sometimes be misled, depending on the source and context.
    \item Score 7-8: High trust, these personas are easily misled and usually do not verify the information.
    \item Score 9-10: Very high trust, these personas almost unquestioningly believe and spread fake news.
\end{itemize}
In each round, you should consider the event and human players' output to evaluate. 
Consider more like a human evaluator, giving more diversity score distribution.

\subsection{Game Mechanics Design}
\label{Game Mechanics Design}
\textbf{Scoring System}: In each round, each LLM character evaluates the messages provided by the players and provides an opinion and trust level score on a 10-point scale:
\begin{itemize}
    \item 10: The character fully trusts the misinformation published by the Influencer.
    \item 0: The character completely trusts the debunking response published by the Journalist.
\end{itemize}
The average score across all five personas determines the outcome of the round. 
The final winning condition is determined by the trust level score generated in the last round. Each round's score is influenced by previous rounds, meaning the score achieved after the final round reflects cumulative performance rather than just performance in the final round:
\begin{itemize}
    \item If the final score is above 5, Influencer wins.
    \item If the final score is below 5, Journalist wins.
    \item If the score equals 5 (indicating neutral public opinion), the player with more remaining in-game currency wins.
\end{itemize}

% \begin{acks}
% thanks.
% \end{acks}
\bibliographystyle{ACM-Reference-Format}
\bibliography{Reference}

\end{document}